\documentclass[aps,prb,twocolumn,letterpaper,showpacs,10pt]{revtex4-1}

\usepackage{graphicx, hyperref}
\usepackage[dvipsnames]{xcolor}
\usepackage{amsmath,mathtools}
\usepackage{amsfonts}
\usepackage{bbm}
\usepackage[mathscr]{euscript}
\usepackage{multirow}

\def\beq{\begin{equation}}
\def\eeq{\end{equation}}
\def\beqn{\begin{eqnarray}}
\def\eeqn{\end{eqnarray}}

\def\R{{\bf R}}

\def\k{{\bf k}}
\def\r{{\bf r}}

\def\u{{\bf u}}

\def\p{{\bf p}}
\def\q{{\bf q}}

\def\0{{\bf 0}}

\def\ket#1{\vert #1 \rangle}
\def\bra#1{\langle #1 \vert}
\def\ev#1{\langle #1 \rangle}
\def\ip#1#2{\langle #1 \vert #2 \rangle}
\def\op#1#2{\vert #1 \rangle \langle #2 \vert}
\def\oev#1#2#3{\langle #1 \vert #2 \vert #3 \rangle}
\renewcommand{\bf}{\mathbf}

\newcommand{\citationtitle}[1]{{\color{Gray}\small #1}}

\begin{document}
\title{The Neutral Excitations in the Gaffnian State}

\author{Byungmin Kang}
\affiliation{Department of Physics, University of California, Berkeley, Berkeley, California 94720, USA}

\author{Joel E. Moore}
\affiliation{Department of Physics, University of California, Berkeley, Berkeley, California 94720, USA}
\affiliation{Materials Sciences Division, Lawrence Berkeley National Laboratory, Berkeley, California 94720, USA}

\date{\today}

\begin{abstract}
We study a model fractional quantum Hall (FQH) wavefunction called the Gaffnian state, which is believed to represent a gapless, strongly correlated state that is very different from conventional metals. To understand this exotic gapless state better, we provide a representation based on work of Halperin in which the pairing structure of the Gaffnian state becomes more explicit. We employ the single-mode approximation introduced by Girvin, MacDonald, and Platzman (GMP), here extended to three-body interactions, in order to treat a neutral collective exitation mode in order to clarify the physical origin of the gaplessness of the Gaffnian state.  We discuss approaches to extract systematically the relevant physics in the long-distance, large-electron-number limit of FQH states using numerical calculations with relatively few electrons.  In an appendix, we provide second quantized expressions for many-body Haldane pseudopotentials in various geometries including the plane, sphere, cylinder, and the torus based on the proper definition of the relative angular momentum.
\end{abstract}


\maketitle

\section{Introduction}
The fractional quantum Hall effect (FQHE) has been a topic of deep and continuing interest since its discovery in 1982~\cite{tsui1982two}. Placed in a strong magnetic field, a two-dimensional electron gas (2DEG) exhibits an exotic order that cannot be described by the conventional Ginzburg-Landau symmetry breaking paradigm.  Incompressible states are characterized by different kinds of topological order~\cite{wen1995topological} created by strong electron-electron interactions. Solving this strongly interacting many-body quantum problem is a daunting task in general, but the model wavefunction approach in many cases successfully captures all the universal features of a many-body electron state in the FQHE.

The model wavefunction approach amounts to simply writing down a trial ground state wavefunction, which, although not the true ground state of the Coulomb interaction Hamiltonian, may have the same topological order as the true ground state. The Laughlin state~\cite{laughlin1983anomalous} and the composite fermion (CF) states~\cite{jainbook,haxton2016} are standard examples of model wavefunctions which are of experimentally relevance.  Beside these \textit{abelian} states, \textit{nonabelian} states including the Pfaffian state and the Read-Rezayi states appear as model wavefunctions in the FQHE.  Non-abelian states can serve as platforms for topological quantum computation~\cite{nayak2008non}. A model wavefunction is often identified as a conformal block~\cite{moore1991nonabelions} of a conformal field theory (CFT)~\cite{cftyellowbook}. The CFT associated with the bulk wavefunction also describes the edge excitations via the bulk/edge correspondence~\cite{dubail2012bulkedge}, and the braiding statistics~\cite{moore1991nonabelions, bonderson2011plasma} among quasi-particles. Even though the model wavefunctions are not eigenstates of the Coulomb interaction, in many cases, there exist \textit{parent} Hamiltonian based on the Haldane pseudopotential approach~\cite{haldane1983sphere, simon2007pseudopotentials}. 

In this paper, we focus on a model wavefunction called the Gaffnian state~\cite{simon2007construction}. Unlike other model FQH states, the Gaffnian state is believed to represent a quantum critical point~\cite{simon2007construction, jolicoeur2014absence, freedman2012galois}, or possibly a gapless phase, rather than an incompressible (gapped) state. Our interest in the Gaffnian state is as a likely example of gapless matter with some remnant of topological physics, even though precisely what ``topological'' means in a gapless state is difficult to define. Unlike the $\nu=1/2$ state~\cite{halperinleeread,sondirac} in which there has recently been a resurgence of interest, the Gaffnian does not currently have an interpretation as a metal of composite fermions. The Gaffnian state can be written as a conformal block of a minimal model CFT $\mathcal{M}(5,3)$, which is a \textit{nonunitary} CFT, unlike the unitary CFTs corresponding to known incompressible states.  Like the Pfaffian state, it is the exact ground state of a parent Hamiltonian involving three-body interactions. In the bosonic case, the parent Hamiltonian is a sum of projectors of the relative angular momentum of three particles onto $\ell=0$ and $\ell=2$:
\begin{equation}
H = A P_3^0 + B P_3^2
\end{equation}
with constants $A,B>0$.   We examine possible gap closing in the neutral sector as we approach the Gaffnian from the nearby composite fermion state with the same filling. In order to look for an excitation becoming gapless in the thermodynamic limit, we employ the single-mode approximation (SMA)~\cite{girvin1986magneto, repellin2014singlemode} by Girvin, MacDonald, and Platzman to test gaplessness of the Gaffnian state.

The paper is organized as follows. In Sec. II, we introduce the Gaffnian state and show it is equivalent to a Halperin-type paired state, which is perhaps surprising as there is no obvious pairing in the standard wavefunction of the Gaffnian.  In Sec. III, we consider the neutral excitations of the Gaffnian state via the single-mode approximation which has a well-defined thermodynamic limit. We also show how to compute density correlation functions efficiently given the second quantized wavefunction of a few electrons.  A torus geometry is used to compute the SMA gap function of the Gaffnian state in order to minimize finite-size effects.  We conclude in Sec. IV with a brief summary of our results and possible future extensions.  Most of the technical aspects are reviewed in detail in the Appendix, which starts by fixing our conventions for physics in the lowest Landau level (LLL) and guiding-center variables.  We then construct many-body pseudopotentials in a plane, cylinder, torus, and sphere using the guiding-center second quantized language only. We also give the more familiar coordinate and momentum space representation of the pseudopotentials.  We end with formulas for the the SMA for three-body interactions that are relevant for the Gaffnian state.

\section{The mode wavefunctions}
A planar lowest-Landau-level wavefunction for $N_e$ particles, which we will call electrons, in the FQHE can be written as
\begin{equation}
\Psi_{N_e} (\r_1,\dots,\r_{N_e}) = P(z_1,\dots,z_{N_e}) e^{-\sum_i |z_i|^2/4},
\label{many_body_wavefunction}
\end{equation}
where $z_j = (x_j+i y_j)/l_B$ is the dimensionless complex coordinate of particle $j$, $l_B=\sqrt{\hbar c/eB}$ is the magnetic length, and $P$ is a (anti-)symmetric polynomial if the electrons are identical bosons (fermions). For simplicity, we will often drop the exponential factor in Eq.~(\ref{many_body_wavefunction}) and only specify the polynomial part of the wavefunction.

In the following, we review the model wavefunction called the Gaffnian state~\cite{simon2007construction} and see how can it be identified as a paired wavefunction of the Halperin type~\cite{halperin1983theory, morf1986microscopic}. The (bosonic) Gaffnian state is the unique highest-density zero energy ground state of a three-body interaction $A P_3^0 + B P_3^2$ ($A,B>0$). $P_3^0$ and $P_3^2$ are examples of Haldane's three-body pseudopotentials~\cite{simon2007pseudopotentials, haldane1983sphere}, and they are projectors of three-body relative angular momentum 0 and 2. The precise definition and explicit expressions can be found in Appendix~\ref{appendix_pseudopotentials}. The Gaffnian state~\cite{simon2007construction} is given by
\begin{align}
\Psi_\textrm{Gf} = \tilde{\mathcal{S}} \Bigg[ &\prod_{a<b}^{N_e/2} (z_a-z_b)^{2+q} \prod_{\frac{N_e}{2} < c<d}^{N_e} (z_c-z_d)^{2+q} \nonumber \\
&\times \prod_{e \le \frac{N_e}{2} < f}^{N_e} (z_e-z_f)^{1+q} \prod_{g=1}^{N_e/2} \frac{1}{(z_g-z_{g+N_e/2})} \Bigg],
\label{gaffnian_wavefunction}
\end{align}
where $q=0$ ($q=1$) and $\tilde{\mathcal{S}}$ is a (anti-)symmetrization among electron indices for bosonic (fermionic) electrons. The Gaffnian state, unlike the Pfaffian~\cite{moore1991nonabelions} and the Haffnian state, cannot be understood as a BCS type paired state~\cite{greiter1992paired, read2000paired} of composite fermions. To see this, suppose that the (fermionic) Gaffnian wavefunction is a BCS-type paired state. Then the wavefunction can be written as~\cite{moller2008paired}
\begin{equation}
\Psi_\textrm{Gf} = \textrm{Pf} [g(\r_i-\r_j)] \prod_{i<j} (z_i-z_j)^2,
\end{equation}
where Pf stands for the Pfaffian of a matrix and $g(\r)$ is an antisymmetric function in $z$. (In the bosonic case, use $(z_i-z_j)$ instead of $(z_i-z_j)^2$.) Let's assume $g(\r)$ has an asymptotic behavior $g(\r) \to C z^\alpha$ as $\r \to \0$, where $C$ is some constant and $\alpha$ can be negative. After factoring out the $\nu=1/2$ Laughlin state (the Jastrow factor squared) from the wavefunction, let us bring $z_3$ to $z_4$, $z_5$ to $z_6$, until $z_{N_e-1}$ to $z_{N_e}$. Finally, the (factored) wavefunction has an asymptotic $\textrm{Pf} [g(\r_i-\r_j)] \to g(\r_1-\r_2) g(\r_3-\r_4) \cdots g(\r_{N_e-1}-\r_{N_e}) \sim (z_3-z_4)^\alpha \cdots (z_{N_e-1} -z_{N_e})^\alpha g(\r_1-\r_2)$. In contrast, the same clustering among particles in the Gaffnian state gives the asymptotic $(z_3-z_4)^{-1} \dots (z_{N_e-1}-z_{N_e})^{-1} \big( \frac{1}{z_1-z_2} \prod_{a=2}^{N_e/2}(z_1-z_{2a-1})(z_2-z_{2a-1}) \prod_{b=a+1}^{N_e/2} (z_{2a-1}-z_{2b-1})^2 \big)$, which cannot be cast into the form of a paired wavefunction of the BCS type.

Hence the nature of the Gaffnian state is very different from other paired FQH states, and it is worthwhile to find an equivalent form of the wavefunction in which the pairing structure of the Gaffnian is more explicit than in Eq.~(\ref{gaffnian_wavefunction}). After a bit of algebra, one can first see that the Gaffnian wavefunction in Eq.~(\ref{gaffnian_wavefunction}) is equal to the permanent wavefunction in Ref.~\onlinecite{yoshioka1988connection}, as was pointed out in the previous literature, e.g., in Ref.~\onlinecite{jolicoeur2014absence}.
\begin{align}
\Psi_\textrm{Gf} = &\prod_{i<j} (z_i-z_j)^{1+q} \nonumber \\
&\times \tilde{\mathcal{S}} \Bigg[ \prod_{a<b}^{N_e/2} (z_a-z_b) \prod_{\frac{N_e}{2} < c<d}^{N_e} (z_c-z_d) \textrm{perm}[M] \Bigg], \nonumber
\end{align}
where $M_{i,j} = (z_i-z_{j+N_e/2})^{-1}$ is the $N_e/2$ by $N_e/2$ matrix and $\textrm{perm}[A] = \sum_{\sigma \in S_n} \prod_{k=1}^{n} A_{k,\sigma(k)}$ is the permanent of \textcolor{red}{a} n-by-n matrix $A$.

This permanent wavefunction was early on used for a trial wavefunction for $\nu=2/5$ state~\cite{fano1986configuration}. In fact, Halperin had earlier suggested a seemingly different trial wavefunction for $\nu=2/5$ state based on the idea of pairing. In a well-known Helv. Phys. Acta. paper~\cite{halperin1983theory}, he presented three different kinds of model wavefunctions - the Halperin paired wavefunction, hierarchy wavefunction, and bilayer wavefunction.  Strictly speaking, the paired wavefunction presented in Ref.~\onlinecite{halperin1983theory} is not a scalar under rotations in the sphere. The scalar version of this wavefunction, which is invariant under the rotations of the sphere, first appeared in Ref.~\onlinecite{morf1986microscopic} slightly before the introduction of the permanent wavefunction~\cite{yoshioka1988connection}, and we will now show that these are actually the same.

To compare the Gaffnian wavefunction and the Halperin paired state, let's define two classes of wavefunctions - ``model paired'' (MP) state and the ``Halperin paired'' (HP) state. We present only the bosonic case in the following.  The corresponding fermionic wavefunction follows after multiplying a Jastrow factor by the bosonic wavefunction.  The model paired (MP) state is defined as
\begin{align}
\Psi_\textrm{MP}^{(p,q,r)} &= \mathcal{S}\bigg[\prod_{1\le a<b}^{N_e/2} (z_a-z_b)^p \prod_{N_e/2 < c<d}^{N_e} (z_c-z_d)^p \nonumber \\
&\times \prod_{a,c =1}^{N_e/2} (z_a-z_{c+N_e/2})^q \prod_{g=1}^{N_e/2} (z_g-z_{g+N_e/2})^{-r} \bigg],
\label{MP}
\end{align}
where $p,q,r$ are positive integers satisfying $q \ge r$ and $\mathcal{S}$ is the symmetrization over electron indices. By counting the total power in $z_i$, we get $N_\phi = \big( \frac{p}{2}+\frac{q}{2} \big) N_e - (p+r)$ which implies the filling factor $\nu=2/(p+q)$ and the shift $\mathscr{S}=p+r$. The MP state is a proper generalization of the Gaffnian wavefunction where the Gaffnian state corresponds to $(p,q,r) = (2,1,1)$. It is also worth mentioning that $(p,q,r) = (2,0,0)$, the complete symmetrization of the $(2,2,0)$ bilayer state, corresponds to the Pfaffian state~\cite{jainbook, papic2010p}. The Halperin paired (HP) state can be defined as
\begin{align}
\Psi_\textrm{HP}^{(s,t,u)} = &\mathcal{S}\bigg[\prod_{i<j}^{N_e} (z_i-z_j)^s \prod_{n=1}^{N_e/2} (z_{2n-1}-z_{2n})^{-t}
\nonumber \\ 
&\times \prod_{1 \le n < m}^{N_e/2} (z_{2n-1}z_{2n}+z_{2m-1}z_{2m}-2Z_n Z_m)^u \bigg],
\label{HP}
\end{align}
where $s,t,u$ are positive integers satisfying $s\ge t$ and $Z_n = (z_{2n-1}+z_{2n})/2$ is the center of mass coordinate between the electron pairs $(z_{2n-1},z_{2n})$. The last term in Eq.~(\ref{HP}) is motivated by the Laughlin state between pairs, i.e. $(Z_n-Z_m)^{2u}$. The last term in Eq.~(\ref{HP}) can also be written as $\frac{1}{2}\big( (z_{2n-1} - z_{2m-1}) (z_{2n}-z_{2m}) + (z_{2n-1}-z_{2m}) (z_{2n}-z_{2m-1}) \big)$. The Halperin paired state has the filling factor $\nu=2/(2s+u)$ and the shift $\mathscr{S}=s+t+u$. At $(s,t,u) = (1,1,1)$~\cite{morf1986microscopic}, the state has the same filling factor and the shift as the Gaffnian state. In the following, we show that the HP state at $(s,t,u) = (1,1,1)$ is indeed {\it equal} to the Gaffnian state; to the best of our knowledge, this equivalence has not been noted in the literature before.

To show $\Psi_\textrm{HP}^{(1,1,1)}$ is the Gaffnian state, it suffices to show that (a) the wavefunction has the same filling factor and shift as the Gaffnian state, (b) it does not vanish as we cluster two particles to the same point, but (c) it vanishes as three or more powers as we cluster three particles to the same point~\cite{simon2007construction,thankstosimon}. The property (a) follows immediately. (b) can be proved by showing the wavefunction does not vanish at $(z_1,z_2,\dots,z_{N_e}) = (w_1,w_1,w_2,w_2,\dots,w_{N_e/2},w_{N_e/2})$, i.e., the wavefunction does not vanish even after clustering all the pairs.  Because of the first term in the RHS of Eq.~(\ref{HP}), most of the terms vanishes when we permute the electron indices. However, there are non vanishing terms, which are all identical and equal to $\prod_{1\le a < b}^{N_e/2} (w_a-w_b)^6$ which implies the condition (b).

Finally, when we cluster three particles to the same point, the first two terms in Eq.~(\ref{HP}) together vanish at least 2 powers and the last term vanishes at least 1 power, so the state vanishes at least 3 powers.  We conclude that a Halperin paired state $\Psi_\textrm{HP}^{(1,1,1)}$ is indeed the Gaffnian state. Further equivalences between the Halperin paired states and the model paired states can be found and are summarized in TABLE~\ref{tableHPeqMP}. It is very surprising that the simple idea of pairing leads to so many exotic \textit{nonabelian} FQH states. Similar results are pointed out recently in Ref.~\onlinecite{jeong2015bilayer}, \onlinecite{repellin2015projective}, and \onlinecite{jeong2016competing}. In Ref.~\onlinecite{jeong2015bilayer} and \onlinecite{repellin2015projective}, equivalences of the Read-Rezayi states and the model paired (MP) states in the torus geometry were found, and in Ref.~\onlinecite{jeong2016competing}, equivalence between the Haffnian state and $\Psi_\textrm{MP}^{(4,0,0)}$ was found, which adds one more entry in TABLE~\ref{tableHPeqMP}.

\begin{table}[t]
\caption{Equivalence between $\Psi_\textrm{HP}^{(s,t,u)}$ and $\Psi_\textrm{MP}^{(p,q,r)}$.}
\begin{tabular}{c|c|c}
FQH state&$\Psi_\textrm{HP}^{(s,t,u)}$&$\Psi_\textrm{MP}^{(p,q,r)}$\\
\toprule
\multirow{2}*{Pfaffian}&\multirow{2}*{(1,1,0)}  &  (2,0,0)\\ & & (1,1,1) \\ \hline
\multirow{2}*{Gaffnian}&\multirow{2}*{(1,1,1)}  &  (2,1,1)\\ & & (1,2,2) \\ \hline
\multirow{2}*{Haffnian}&\multirow{2}*{(1,1,2)}  &  (2,2,2)\\ & & (1,3,3)
\end{tabular}
\label{tableHPeqMP}
\end{table}

\section{Neutral excitations in the Gaffnian state}
Another description of the Gaffnian wavefunction~\cite{simon2007construction} is a conformal block of a conformal field theory~\cite{cftyellowbook}. The Gaffnian wavefunction can be expressed as
\begin{equation}
\Psi_\textrm{Gf} = \oev{0}{O(z_1) \cdots O(z_{N_e}) O_\textrm{bg}}{0},
\end{equation}
where $O(z) = \psi(z) e^{i \phi_c (z)/\sqrt{\nu}}$ with $\phi_c$ being the free boson of a $U(1)$ CFT and $\psi(z)$ being a field with scaling dimension $\Delta_\psi = 3/4$ in the minimal model $\mathcal{M}(5,3)$. The background charge operator $O_\textrm{bg}$ is introduced to impose charge neutrality condition so that the conformal block does not vanish. A CFT used to construct a bulk wavefunction also describes the edge excitations, quasi-hole excitations, and the braidings of the quasi-holes~\cite{moore1991nonabelions, simon2007construction, bonderson2011plasma}. This is a manifestation of the bulk/edge correspondence~\cite{dubail2012bulkedge}, which requires a bulk \textit{gap} in order to have well-defined \textit{gapless} edge and quasi-hole excitations constructed from the CFT.

The CFT for the Gaffnian state is \textit{nonunitary} resulting in a nonunitary braiding among quasi-holes~\cite{wu2014braiding}, which is not physically sensible. There has been an argument~\cite{read2009conformal}, numerical evidence in the sphere geometry~\cite{jolicoeur2014absence}, and even a proof~\cite{freedman2012galois} that the Gaffnian state represents a gapless state, but the physical picture of how the Gaffnian state becomes gapless remains elusive so far. To understand its gaplessness better, we focus on the neutral excitations of the Gaffnian state and examine the physical mechanism of closing of the gap.

The neutral excitations of a FQH state are collective excitations of quasi-electron and quasi-hole pairs.  In an incompressible state, there is typically a well-defined low-lying neutral excitation mode, which is called the magnetoroton mode~\cite{haldane1985finite, girvin1986magneto, repellin2014singlemode}.  In a seminal paper~\cite{girvin1986magentoroton}, Girvin, MacDonald, and Platzman (GMP) modeled the magnetoroton mode by the single-mode approximation (SMA) which is the (projected) density-wave excitation above the ground state. This trial excitation wavefunction correctly captures the qualitative and quantitative behavior of the magnetoroton mode, especially the location and the size of the gap~\cite{girvin1986magentoroton, repellin2014singlemode}.

It should be noted that (a) the Gaffnian state is known to become gapped in a thin cylinder~\cite{papic2014solvable, weerasinghe2014thin}, and (b) the $\nu=2/3$ (bosonic) Jain composite fermion (CF) state~\cite{jainbook, haxton2016} is a nearby phase of the Gaffnian state, as the Jain state is an incompressible ground sate of a pseudopotential $P_2^0$ with the same filling factor and shift as the Gaffnian. Starting from the Hamiltonian $A P_3^0 + B P_3^2$ ($A,B>0$) for which the Gaffnian state is the ground state, one can perturb the Hamiltonian by $P_2^0$ so that the ground state belongs to the same universality class as the CF state after the gap opens. Conversely, we can adiabatically change the Hamiltonian from the CF state to the Gaffnian state and close the gap along the way. Together with the gapped nature in the thin cylinder geometry, the Gaffnian state seems to represent a \textit{quantum critical point} with a neighboring CF state. 

Guided by a recent Exact Diagonalization (ED) study~\cite{jolicoeur2014absence}, we argue that the gap closing happens at the wave vector $\q=\0$ at the Gaffnian state. (This kind of gap closing scenario has been studied before~\cite{scarola2000excitonic, mulligan2010isotropic, you2014theory} in the FQHE in various other situations.) To this end, we employ the SMA~\cite{girvin1986magentoroton, repellin2014singlemode} of the magnetoroton mode to capture the possible gap closing nature of the Gaffnian state. Even though the magnetoroton mode is not very well-defined in the non-abelian FQH states as it generally contains several low-lying excitation modes~\cite{repellin2015projective}, the SMA can still serve as a trial model or upper bound for the low-lying excitation modes. Also, it was pointed out recently~\cite{repellin2014singlemode} that the SMA becomes a better and better approximation as we approach to the $\0$ wave vector. So SMA provides a way to test the scenario of a gap to neutral excitations closing at $\q=\0$. In the following, we review the basics of the SMA and provide the full expressions generalizing it for a three-body interaction, with details relegated to Appendix~\ref{appendix_SMA}.

\subsection{Single-mode approximation}
The single-mode approximation (SMA) is constructed by applying the guiding-center (projected) density operator to the ground state $\ket\Psi$:
\begin{equation}
\ket{\Psi_\q^\textrm{SMA}} = \frac{1}{\sqrt{N_e}} \hat\rho_\q^\dagger \ket\Psi,
\label{SMA_wavefunction}
\end{equation}
where $\hat\rho_\q = \sum_i e^{-\q \cdot \R_i}$ is the guiding-center density operator with $\R_i = (X_i,Y_i)$ being the guiding-center coordinate of particle $i$, and $\q \ne \0$. The guiding-center density operator satisfies nontrivial commutation relations called the GMP or magnetic translation algebra~\cite{girvin1986magentoroton, murthy2003review, bernevig2012emergent}: 
\begin{equation}
[\hat\rho_{\q_1} , \hat\rho_{\q_2}] = 2i \sin \Big( \frac{1}{2} l_B^2 \q_1 \wedge \q_2 \Big) \hat\rho_{\q_1+\q_2}
\end{equation}
The SMA is orthogonal to the ground state (for $\q \ne \0$) if the ground state $\ket\Psi$ is homogeneous, i.e., it has a constant one-particle density~\cite{FootnoteSMA}. Using Eq.~(\ref{SMA_wavefunction}), the gap function is given by~\cite{girvin1986magentoroton}
\begin{equation}
\Delta(\q) = \frac{ \oev{\Psi_\q^\textrm{SMA}}{(H - E_\textrm{GS})}{\Psi_\q^\textrm{SMA}} }{ \ip{\Psi_\q^\textrm{SMA}}{\Psi_\q^\textrm{SMA}} } = \frac{\hat{f}(\q)}{\hat{s}(\q)},
\label{gap_function}
\end{equation}
where $\hat{s}(\q)$ is the (guiding-center) structure factor, $\hat{f}(\q) = \frac{1}{2 N_e} \oev{\Psi_\q^\textrm{SMA}}{\big[ \hat\rho_\q, \big[\overline{H}, \hat\rho_\q^\dagger\big] \big]}{\Psi_\q^\textrm{SMA}}$, and $H$ is the interaction Hamiltonian. (The kinetic part of the Hamiltonian is quenched since we are only interested in the states in the LLL. $\overline{\mathcal{O}}$ denotes the LLL projection of an operator $\mathcal{O}$.)

The Gaffnian state is a zero energy ground state of a three-body interaction
\begin{align}
V(\k_1,\k_2) &= A (2\pi)^2 + B (2\pi)^2 \bigg( L_2 \Big( \frac{l_1^2 |\k_1|^2}{2} \Big) \nonumber \\
&  + L_1 \Big( \frac{l_1^2 |\k_1|^2}{2} \Big) L_1 \Big( \frac{l_2^2 |\k_2|^2}{2} \Big) + L_2 \Big( \frac{l_2^2 |\k_2|^2}{2} \Big) \bigg), 
\label{Gaffnian_hamiltonian_momentum}
\end{align}
where $A,B>0$, $L_m(x)$ is the $m$-th Laguerre polynomial, and $(l_1, l_2) = \big(\sqrt{2} l_B, \sqrt{\frac{3}{2}} l_B \big)$. The Jacobi-coordinate system used in Eq.~(\ref{Gaffnian_hamiltonian_momentum}), which is explained in the Appendix.

In the case of two-body interaction, $\hat{f}(\q)$ in the gap function Eq.~(\ref{gap_function}) is given by~\cite{girvin1986magentoroton}
\begin{align}
\hat{f}(\q) = \int \frac{d^2 (l_1 \k)}{(2\pi)^2} \Big( v(&\k+\q) - 2 v(\k) + v(\k-\q) \Big) \nonumber \\
&\times 2 \sin^2 \Big( \frac{1}{2} l_B^2 \q \wedge \k \Big) \hat{s}(\k),
\end{align}
where $v(\k) = V(\k) e^{-\frac{1}{4} l_1^2 |\k|^2}$ with $V(\k)$ being the Fourier transformation of a two-body interaction $V(\r_1-\r_2)$, and $\hat{s}(\k)$ is the (guiding-center) structure factor. So the gap function depends only on (a) the interaction Hamiltonian and (b) the structure factor of the ground state, which encodes essential information of the state in the thermodynamic limit. Generalization to the three-body interaction is straightforward but requires lengthy algebra. Rather than going into these cumbersome steps, which are done in detail in Appendix~\ref{appendix_SMA}, we simply mention that for three-body interaction we need one additional ingredient to compute the gap function: (c) the three-density correlation function. The density correlation functions are interesting in their own right as they contain information about the state in the thermodynamic limit.

\subsection{Density correlation functions}
In this section, we review density correlation functions and discuss how one can compute them from the state written in terms of a (occupation number) second quantized basis. We work in the plane and use the symmetric gauge. The simplest example of a density correlation function is the one-particle density, which measures the probability of finding a particle at position $\r$:
\begin{align}
\rho (\r) &= \ev{\hat\psi^\dagger (\r) \hat\psi (\r)} \nonumber \\
&= N_e \int d^2 \r_2 \dots d^2 \r_{N_e} |\Psi(\r,\r_2,\dots,\r_{N_e})|^2,
\end{align}
where $\hat\psi(\r) = \sum_{m} \phi_m (\r) c_m$ is the annihilation operator at $\r$, $\phi_m(\r) = \frac{1}{\sqrt{2\pi 2^m m! l_B^2}} z^m e^{-\frac{1}{4} |z|^2}$ is the normalized wavefunction of orbital $m$ in the symmetric gauge, and $c_m$ is the annihilation operator associated with orbital $m$. The next example is the pair-correlation function:
\begin{align}
g&(\R,\u) = \rho^{-2} \ev{\hat\psi^\dagger (\r_1) \hat \psi^\dagger (\r_2) \hat \psi(\r_2) \hat \psi(\r_1)} \nonumber \\
&= \frac{N_e (N_e-1)}{\rho^2} \int d^2 \r_3 \dots d^2 \r_{N_e} |\Psi(\r_1,\dots,\r_{N_e})|^2,
\label{pair_correlation}
\end{align}
where $\rho = \frac{\nu}{2\pi}$ is the density of the (homogeneous) ground state and $(\R,\u)$ and $(\u_1,\u_2)$ are related by the two-body Jacobi coordinates: $(\r_1,\r_2) = \big( \R + \frac{\u}{2}, \R - \frac{\u}{2} \big)$. The third density correlation function is the three-density correlation function:
\begin{align}
h(\R,&\u_1,\u_2) = \rho^{-3} \ev{\hat\psi^\dagger (\r_1) \hat \psi^\dagger (\r_2) \hat \psi^\dagger (\r_3) \hat \psi(\r_3) \hat \psi(\r_2) \hat \psi(\r_1)} \nonumber \\
&= \frac{N_e (N_e-1) (N_e-2)}{\rho^3} \int d^2 \r_4 \dots d^2 \r_{N_e} |\Psi|^2,
\end{align}
where we have used the three-body Jacobi coordinates: $(\r_1,\r_2,\r_3) = \big(\R+\frac{\u_1}{2}+\frac{\u_2}{3}, \R-\frac{\u_1}{2}+\frac{\u_2}{3}, \R - \frac{2}{3} \u_2 \big)$. In the thermodynamic limit, the state we are mainly interested in becomes homogeneous and all the density correlation functions are independent of the CM coordinate:
\begin{align}
\rho(\r) &\to \rho \nonumber \\
g(\R,\u) &\to g(\u) \nonumber \\
h(\R,\u_1,\u_2) &\to h(\u_1,\u_2) \nonumber
\end{align}
as $N_e \to \infty$. In the case of a three-body interaction, $g(\u)$ and $h(\u_1,\u_2)$ enter in the expression of the SMA gap function.

The Fourier transformation of the pair-correlation functions is called the structure factor:
\begin{equation}
s(\k) = 1+ \int d^2 \u e^{i\k \cdot \u} \rho \big( g(\u)-1 \big).
\label{structure_factor}
\end{equation}
The Fourier transformation of the three-density correlation function is given by:
\begin{widetext}
\begin{align}
\Lambda (\k_1,\k_2) = &-2 + s\Big( \k_1 + \frac{1}{2}\k_2 \Big) + s\Big(-\k_1 + \frac{1}{2}\k_2 \Big) + s(\k_2) \nonumber \\
&+ \int d^2 \u_1 d^2 \u_2 e^{i \k_1 \cdot \u_1} e^{i \k_2 \cdot \u_2} \rho^2 \bigg( h(\u_1,\u_2) - g \Big( \frac{1}{2} \u_1 + \u_2 \Big) - g \Big( -\frac{1}{2} \u_1 + \u_2 \Big) -g(\u_1) +2 \bigg).
\label{FT_three_density}
\end{align}
The guiding-center (or projective) analog of Eq.~(\ref{structure_factor}) and (\ref{FT_three_density}) are given by
\begin{equation}
\hat{s}(\k) = 1+ e^{\frac{1}{4} l_1^2 |\k|^2} \int d^2 \u e^{i\k \cdot \u} \rho \big( g(\u)-1 \big)
\end{equation}
and
\begin{align}
\hat{\Lambda} (\k_1,&\k_2) = - \Big( e^{\frac{i}{2} l_B^2 \k_1 \wedge \k_2} + e^{-\frac{i}{2} l_B^2 \k_1 \wedge \k_2} \Big) +  e^{\frac{i}{2} l_B^2 \k_1 \wedge \k_2} \hat{s}(\k_2) + e^{-\frac{i}{2} l_B^2 \k_1 \wedge \k_2} \hat{s}\Big( \k_1 + \frac{1}{2}\k_2 \Big) + e^{\frac{i}{2} l_B^2 \k_1 \wedge \k_2} \hat{s}\Big( -\k_1 + \frac{1}{2}\k_2 \Big) \nonumber \\
&+ e^{\frac{1}{4} l_1^2 |\k_1|^2} e^{\frac{1}{4} l_2^2 |\k_2|^2} \int d^2 \u_1 \int d^2 \u_2 e^{i\k_1 \cdot \u_1} e^{i\k_2 \cdot \u_2} \rho^2 \bigg( h(\u_1,\u_2) -g\Big(\frac{1}{2}\u_1 +\u_2\Big) -g\Big(-\frac{1}{2}\u_1 +\u_2\Big) -g(\u_1) +2 \bigg),
\label{projected_FT_three_density}
\end{align}
where we have used $\k_1 \wedge \k_2 = (\k_1)_x (\k_2)_y - (\k_1)_y (\k_2)_x$. Because Eq.~(\ref{projected_FT_three_density}) lacks explicit symmetries, we often use the symmetrized version of Eq.~(\ref{projected_FT_three_density}):
\begin{align}
\hat{\Lambda}^\textrm{sym} (\k_1,&\k_2) = \frac{1}{6} \bigg[ \hat{\Lambda}(\k_1,\k_2) + \hat{\Lambda}(-\k_1,\k_2) + \hat{\Lambda}\Big(\frac{1}{2}\k_1 + \frac{3}{4}\k_2,\k_1-\frac{1}{2}\k_2 \Big) + \hat{\Lambda}\Big(-\frac{1}{2}\k_1 + \frac{3}{4}\k_2,-\k_1-\frac{1}{2}\k_2 \Big) \nonumber \\
&\qquad \qquad \qquad \qquad + \hat{\Lambda}\Big(-\frac{1}{2}\k_1 - \frac{3}{4}\k_2,\k_1-\frac{1}{2}\k_2 \Big) + \hat{\Lambda}\Big(\frac{1}{2}\k_1 - \frac{3}{4}\k_2,-\k_1-\frac{1}{2}\k_2 \Big) \bigg] \nonumber \\
&= 2 \cos\Big(\frac{1}{2} l_B^2 \k_1 \wedge \k_2 \Big) \bigg( -1 + \frac{1}{2}\hat{s}(\k_2) + \frac{1}{2}\hat{s}\big( \k_1 + \frac{1}{2}\k_2 \big) + \frac{1}{2}\hat{s}\big( -\k_1 + \frac{1}{2}\k_2 \big) \bigg) \nonumber \\
&\quad + e^{\frac{1}{4} l_1^2 |\k_1|^2} e^{\frac{1}{4} l_2^2 |\k_2|^2} \int d^2 \u_1 d^2 \u_2 e^{i\k_1 \cdot \u_1} e^{i\k_2 \cdot \u_2} \rho^2 \Big( \eta(\u_1,\u_2) -g\big(\frac{1}{2}\u_1 +\u_2\big) -g\big(-\frac{1}{2}\u_1 +\u_2\big) -g(\u_1) +2 \Big).
\end{align}
\end{widetext}
A detailed explanation of density correlation functions and their symmetries is presented in Appendix~\ref{appendix_density_correlation_functions}.

Given the model wavefunction, the conventional method for computing the pair-correlation function (which is a two-density correlation function) is to numerically compute the integral in Eq.~(\ref{pair_correlation}) using the Metropolis Monte Carlo algorithm~\cite{metropolis_MC}. This works well for many model FQH states including the Laughlin state~\cite{girvin1986magentoroton}, the composite fermion states~\cite{jainbook, scarola2000excitonic}, and the Pfaffian state~\cite{park1998possibility} - all of which have an efficient way of computing the wavefunction modulus squared. Because of numerically expensive symmetrization in the Gaffnian state Eq.~(\ref{gaffnian_wavefunction}), the Metropolis Monte Carlo algorithm requires exponential computing time in the number of particle in order to compute the pair-correlation function. Rather than sticking with this conventional method, we present an alternative route to computing the density-correlation functions using the expressions of a state in the second-quantization. It is known that the Gaffnian state can be written as the Jack polynomial~\cite{bernevig2008model, bernevig2008generalized, bernevig2008properties, bernevig2009clustering} for which the second quantized expression can be computed efficiently~\cite{bernevig2009anatomy, thomale2011decomposition}. Given the wavefunction in second quantization (to be more precise, in the occupation number basis), we consider the following combination~\cite{girvin1986magentoroton}:
\begin{align}
\rho \big( g(\u) - 1\big) = \frac{1}{\nu} \sum_{p=0}^{N_\phi} &|\phi_p(\u)|^2 \Big( \ev{(c_p c_0)^\dagger (c_p c_0)} \nonumber \\
&- \ev{c_p^\dagger c_p} \ev{c_0^\dagger c_0} + \big( \ev{c_0^\dagger c_0} -\nu \big) \delta_{p,0} \Big),
\label{finite_pair_correlation}
\end{align}
where $\rho= \frac{\nu}{2\pi}$ is the one-particle density of the system (in the thermodynamic limit) and we are considering the case of a finite number of particles: $N_\phi = \frac{1}{\nu} N_e - \mathscr{S}$. This particular combination imposes the sum rules~\cite{girvin1986magentoroton}:
\begin{align}
&\int d^2 \u \rho \big( g(\u)-1 \big) = -1 \\
&\int d^2 \u \Big( \frac{\u^2}{2} \Big) \rho \big( g(\u)-1 \big) = -1,
\end{align}
which ensures the structure factor has an expansion $s(\k) = \frac{\k^2}{2} + \mathcal{O} (\k^4)$. The three-density correlation function also has a similar expression:
\begin{widetext}
\begin{align}
\rho^2 \bigg( &h(\u_1,\u_2) - g \Big( \frac{1}{2} \u_1 + \u_2 \Big) - g \Big( -\frac{1}{2} \u_1 + \u_2 \Big) -g(\u_1) +2 \bigg) \nonumber \\
&= \frac{1}{\nu} \sum_{\substack{p_1,p_2,q_1,q_2=0 \\ p_1+p_2 = q_1 +q_2}}^{N_\phi} \phi_{p_1} (-\u_1) \phi_{q_1}^* (-\u_1) \phi_{p_2} \Big( -\frac{1}{2}\u_1 - \u_2 \Big) \phi_{q_2}^* \Big( -\frac{1}{2}\u_1 - \u_2 \Big) C_{p_1,p_2;q_1,q_2},
\label{finite_three_density}
\end{align}
where
\begin{align}
C_{p_1,p_2;q_1,q_2} = &\ev{(c_{q_1} c_{q_2} c_0)^\dagger (c_{p_1} c_{p_2} c_0)} - \ev{(c_{q_1} c_{q_2})^\dagger (c_{p_1} c_{p_2})} \ev{c_0^\dagger c_0} \nonumber \\
& + \Big(- \ev{(c_{p_1} c_{0})^\dagger (c_{p_1} c_{0})} \ev{c_{p_2}^\dagger c_{p_2}} - \ev{(c_{p_2} c_{0})^\dagger (c_{p_2} c_{0})} \ev{c_{p_1}^\dagger c_{p_1}} + 2 \ev{c_{p_1}^\dagger c_{p_1}} \ev{c_{p_2}^\dagger c_{p_2}} \ev{c_{0}^\dagger c_{0}} \Big) \delta_{p_1,q_1}.
\end{align}
Eq.~(\ref{finite_three_density}) satisfies the following sum rule:
\begin{equation}
\int d^2 \u_2 \, \rho^2 \bigg( h(\u_1,\u_2) - g \Big( \frac{1}{2} \u_1 + \u_2 \Big) - g \Big( -\frac{1}{2} \u_1 + \u_2 \Big) -g(\u_1) +2 \bigg) = -2 \rho \big( g(\u_1) - 1\big).
\end{equation}
\end{widetext}

\begin{figure}[t]
\includegraphics[width=1.0\columnwidth]{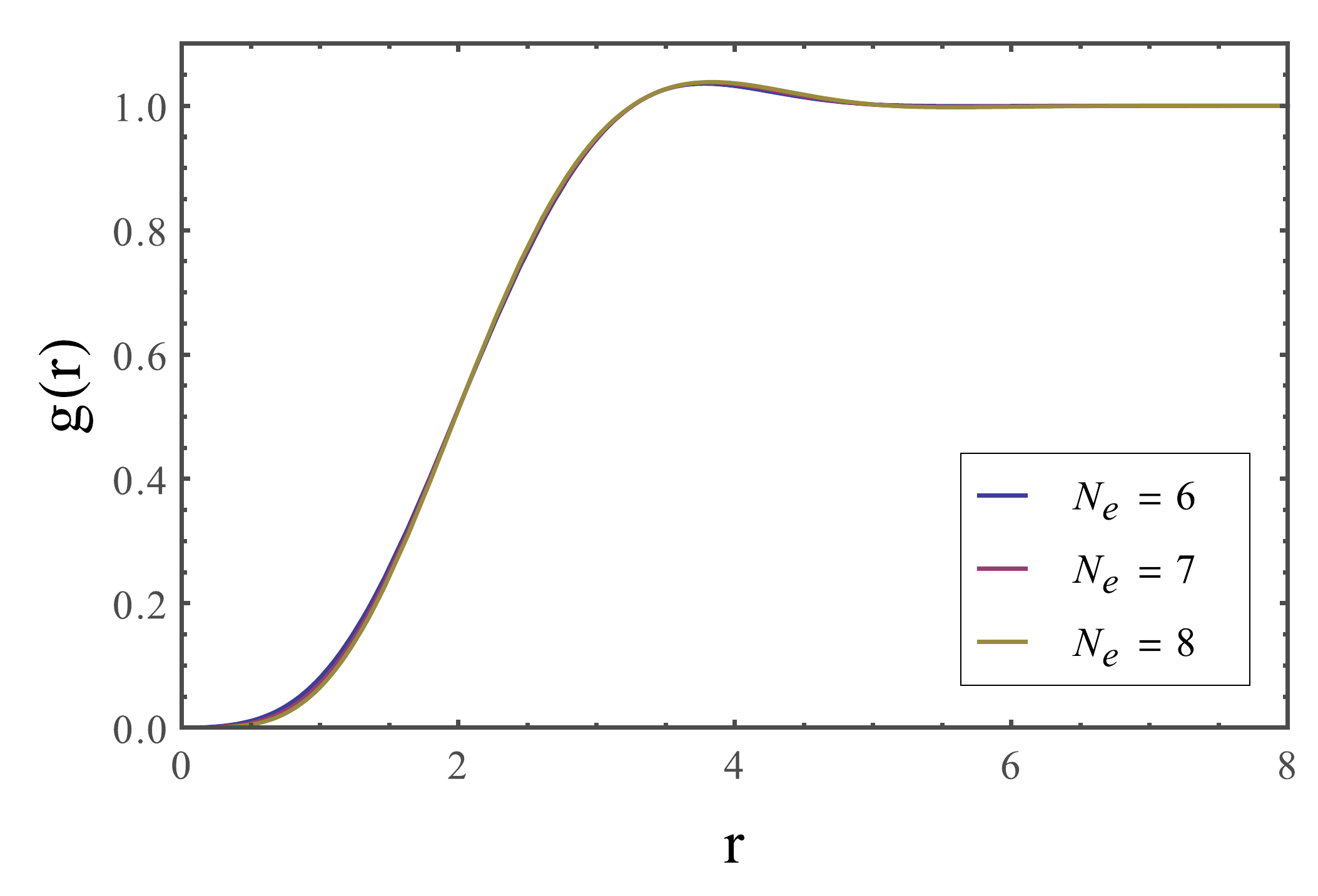}
\caption{The pair correlation function $g(r) = \rho^{-2} \langle{\hat\psi^\dagger({\bf 0}) \hat\psi^\dagger({\bf r}) \hat\psi({\bf r}) \hat\psi({\bf 0})}\rangle$ of the $\nu=1/2$ Laughlin state in a plane for $N_e = 6,7,8$ particles.}
\label{fig_laughlin_pair_correlation}
\end{figure}

\begin{figure}[t]
\includegraphics[width=1.0\columnwidth]{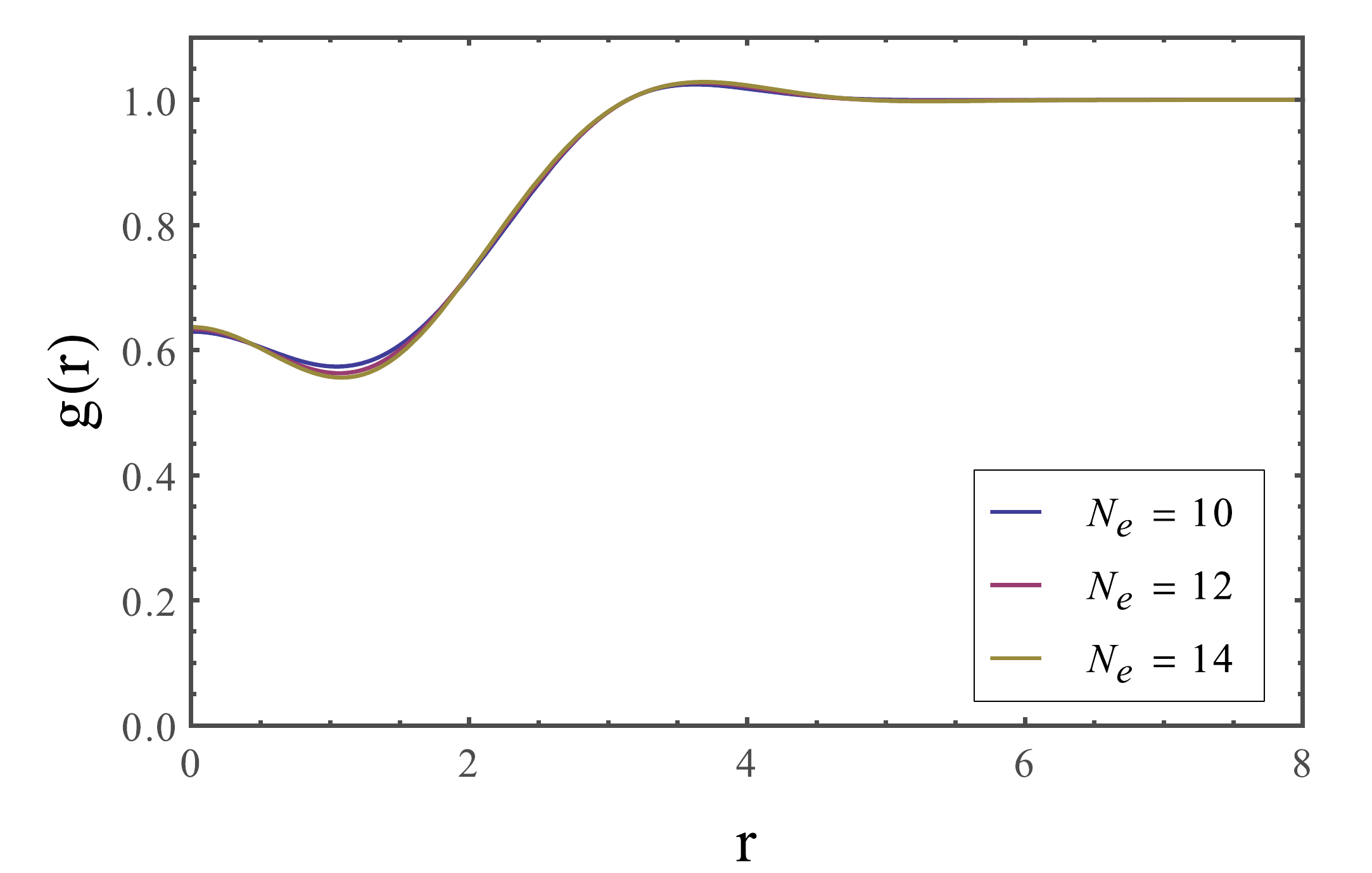}
\caption{The pair correlation function $g(r)$ of the $\nu=1$ Pfaffian state for $N_e = 10,12,14$ particles.}
\label{fig_pfaffian_pair_correlation}
\end{figure}

\begin{figure}[t]
\includegraphics[width=1.0\columnwidth]{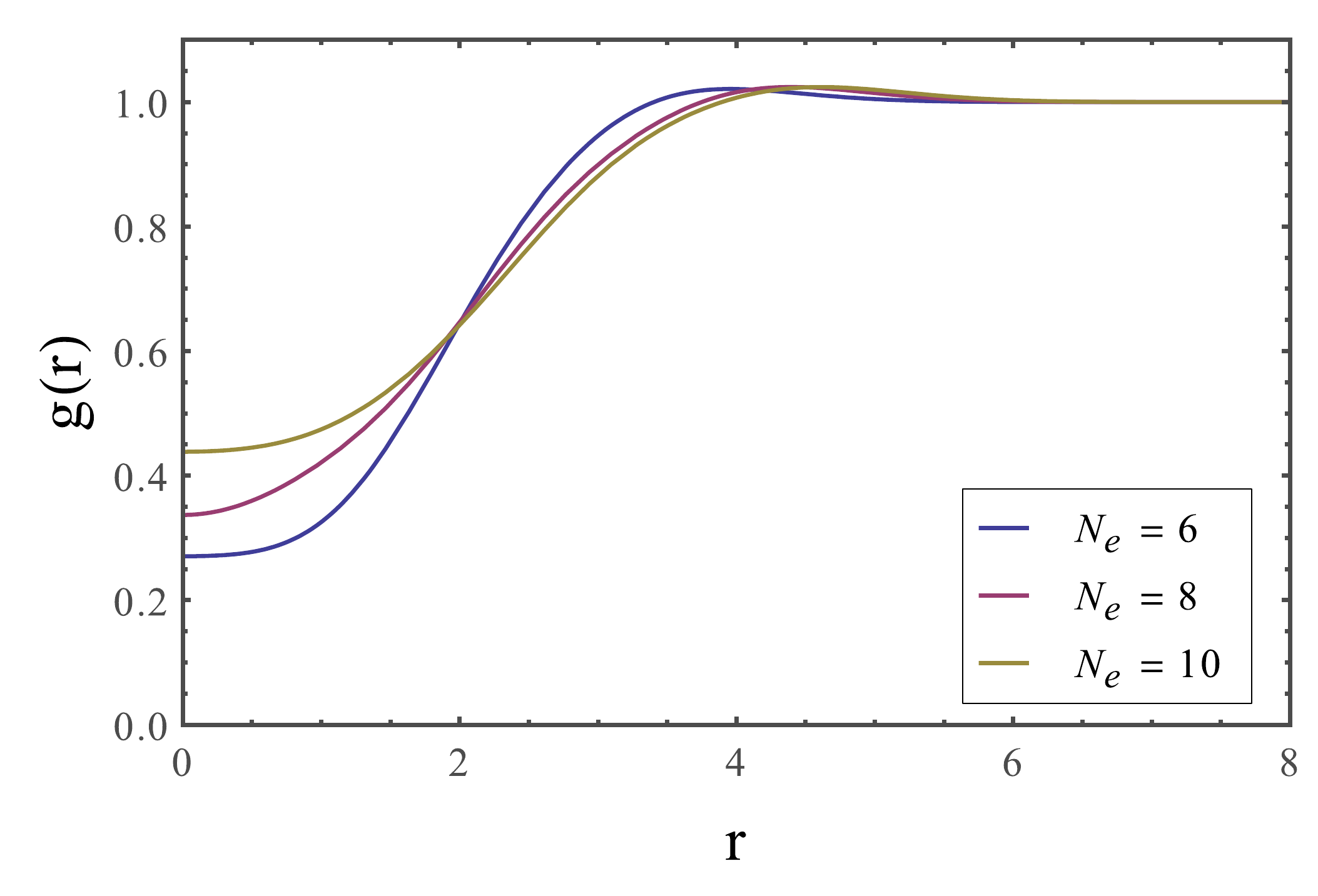}
\caption{The pair correlation function $g(r)$ of the $\nu=2/3$ Gaffnian state for $N_e = 6,8,10$ particles.}
\label{fig_gaffnian_pair_correlation}
\end{figure}

There are several advantages in using second quantized expressions. First of all, it is a \textit{numerically-exact} method. The final expression is the sum of the expectation values of certain operators so an exact computation is possible. Also, the expectation values entering in the expressions can be efficiently evaluated which significantly reduces the computing time. Another advantage is that the result with few particles already well approximates the thermodynamic limit function as shown in Fig.~(\ref{fig_laughlin_pair_correlation}) and (\ref{fig_pfaffian_pair_correlation}). The Monte-Carlo evaluation in a sphere~\cite{caillol1982plasma} also gives a quite good approximation even for a few number of particle, however, it suffers from a finite cut-off, i.e., the pair correlation function ends at finite value of $r$, due to its compact geometry. Also, the evaluation of the three-density correlation function is challenging using Monte Carlo.

\subsection{Gap function of the Gaffnian state}
Using the three-body interaction Eq.~(\ref{Gaffnian_hamiltonian_momentum}) and the two- and three-density correlation functions, the gap function of the SMA of the Gaffnian state can be computed. However, it turns out that the interaction potential in Eq.~(\ref{Gaffnian_hamiltonian_momentum}) is extremely sensitive to the sub-leading terms in the two- and three-density correlation functions which are small in the original correlation functions but largely amplified in the gap function when using Eq.~(\ref{Gaffnian_hamiltonian_momentum}) as an interacting Hamiltonian. To overcome such difficulties and to minimize finite size effects, we employ the torus geometry by following the prescription in Ref.~\onlinecite{repellin2014singlemode}.

The bosonic Gaffnian state (or to be more precise, the Hamiltonian $A P_3^0 + B P_3^2$ with $A,B>0$) has 6-fold degenerate zero energy ground states in the torus geometry. With considerations of the many-particle translational symmetries, the Gaffnian state has 2-fold degeneracy in $(k_x,k_y) = (0,0)$ sector and other ground states are obtained by acting with many-particle translational symmetries to this 2-fold ground states. We lift the 2-fold degeneracy by adding a small perturbation $P_2^0$, the two-body relative angular momentum 0 projector. When the perturbation becomes large, this ground state eventually belongs to the $\nu=2/3$ composite fermion phase. The construction of the Hamiltonian in the torus geometry is extensively reviewed in Appendix.~\ref{appendix_pseudopotentials} and relevant many-particle translational symmetries are well explained in Ref.~\onlinecite{chakrabortyFQHEbook, haldane1985many, bernevig2012emergent}.

\begin{figure}[t]
\includegraphics[width=1.0\columnwidth]{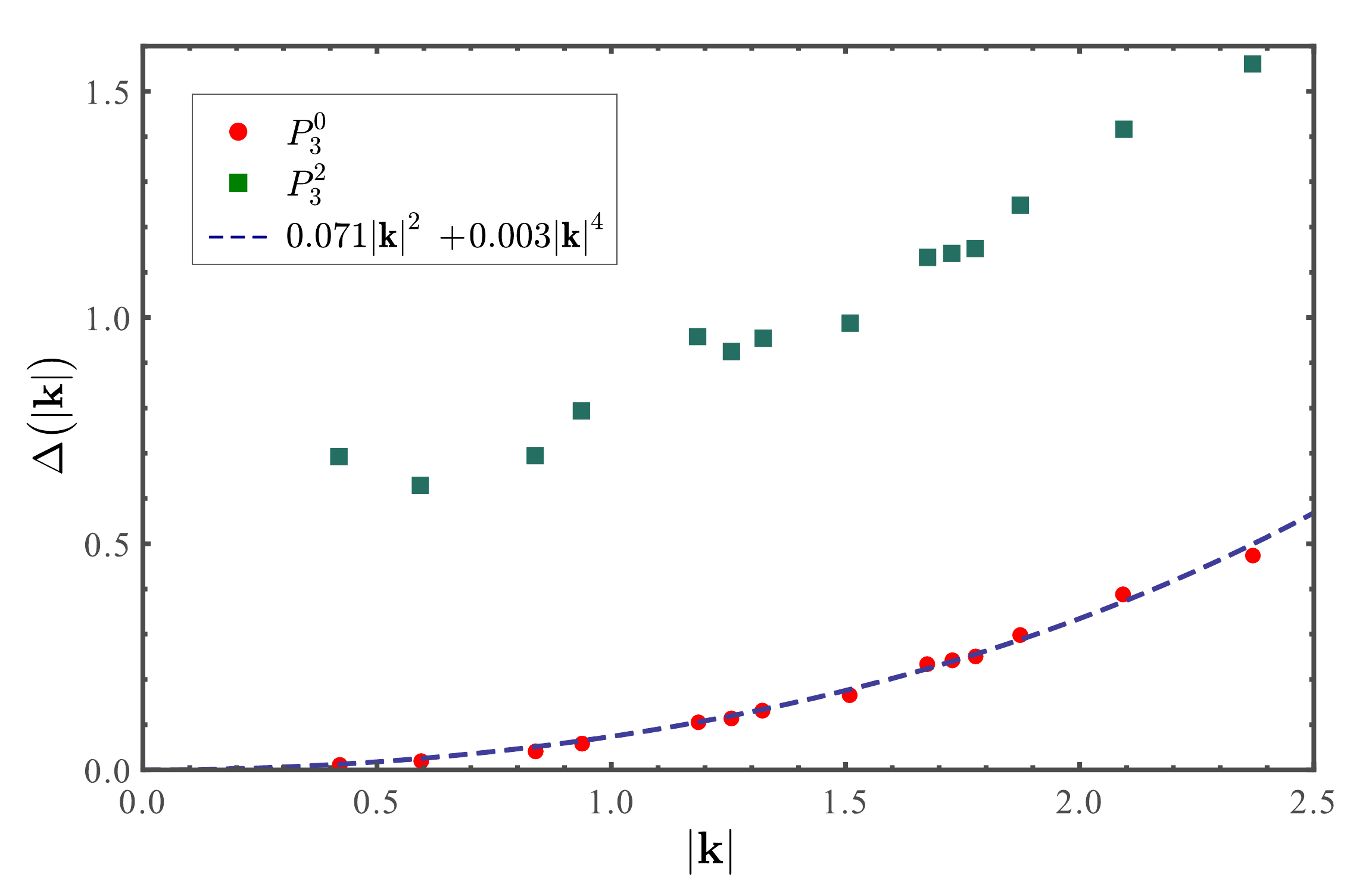}
\caption{The gap function $\Delta\big(|\k|\big)$ of the SMA of the Gaffnian state in the torus v.s. momentum $|\k|$ in a torus with an aspect ratio 1 and zero twisting angle. $(N_e,N_\phi) = (10,15)$ is used to plot the gap function associated with $P_3^0$ and $P_3^2$ using the SMA. $P_3^0$ has a quadratic gap closing at $\k=\0$ with all data points lying in a fitted curve. $P_3^2$ suggests a finite gap at $\k=\0$ with data points not lying in a single curve.}
\label{fig_gaffnian_SMA}
\end{figure}

The numerical result is summarized in Fig.~(\ref{fig_gaffnian_SMA}) in which $N_e=10$ electrons in a torus with an aspect ratio $1$ and zero twisting angle are used. The SMA gives quadratic gap closing at $\k=\0$ under the pseudopotential $P_3^0$, but it seems to give a finite gap for the accessible particle numbers when the pseudopotential $P_3^2$ is nonzero. Although the SMA doesn't give conclusive evidence of the gaplessness of the Gaffnian state, it has its gap minimum at (or around) $\k=\0$.

\section{Conclusion}
In this paper, we have studied an exotic FQH state called the Gaffnian state, which is believed to represent a quantum critical point. We have shown that the pairing structure of the Gaffnian state, which cannot be understood in terms of the BCS pairing, can be understood in terms of a pairing of the Halperin type.  It is also shown that nonabelian states including the Pfaffian and the Haffnian state can also be cast into the Halperin paired state form. As the density correlation functions contain useful information about the state in the thermodynamic limit, we present an efficient way to computing those.  Our method relies on the second quantized expression of the state, which works well for the state for which the second quantized expression can be computed efficiently, including the Gaffnian state.

To test the gap closing scenario at $\k=\0$ of the Gaffnian state as we approach from a nearby composite fermion state, we employed the single-mode approximation to compute the gap function associated with the Gaffnian state. We used a torus geometry, in the hope of reducing finite-size effects, with up to $10$ electrons. This approach did not give conclusive evidence for the gaplessness of the Gaffnian state.  However, our numerical results do seem to support the scenario of a gap closing at $\k=\0$.  We hope that the three-body SMA approach developed here will be useful for the many interesting quantum Hall states, either gapped or gapless, that arise naturally in the presence of three-body interactions.

\begin{acknowledgments}
We thank G. Y. Cho, Y.-M. Lu, K. Park, Y.-J. Park, and S. Simon for illuminating discussions. B.K. and J.E.M. were supported by NSF grant DMR-1507141 and an Investigator grant from the Simons Foundation.
\end{acknowledgments}

\appendix

\section{The lowest Landau level physics in various gauge/geometry}

\subsection{The guiding-center coordinates}
In this Appendix, we introduce the guiding-center degrees of freedom, which are true physical degrees of freedom in the lowest Landau level (LLL). Consider electrons with charge $-e$ ($<0$) moving in a 2D (infinite) plane in a strong magnetic field $-B\hat{z}$ ($B>0$). The Hamiltonian of $N$ electrons is
\begin{eqnarray}
H &=& \frac{1}{2m} \sum_{i=1}^{N} \big(\p_i + \frac{e}{c}\bf{A}_i \big)^2 + V(\r_1,\dots,\r_N) \nonumber \\
&=& \frac{1}{2m} \sum_{i=1}^{N} \boldsymbol{\pi}_i^2 + V(\r_1,\dots,\r_N),
\end{eqnarray}
where $\boldsymbol{\pi} = \p + \frac{e}{c} \bf{A}$ is the dynamical momenta, $\bf{A}$ is the vector potential associated with the magnetic field $-B\hat{z}$, and $V$ is the interaction. Due to the presence of the magnetic field, dynamical momenta have nontrivial commutation relation $[\pi_x,\pi_y] = i \hbar^2/l_B^2$, where $l_B=\sqrt{\frac{\hbar c}{eB}}$ is the magnetic length. Without interaction, the Hamiltonian reduces to the one dimensional quantum harmonic oscillator Hamiltonian. It is then useful to define creation/annihilation operator,
\begin{eqnarray}
a &=& \frac{i}{\sqrt{2}} \frac{\pi_x+i\pi_y}{\hbar/l_B} \nonumber \\
a^\dagger &=& \frac{-i}{\sqrt{2}} \frac{\pi_x-i\pi_y}{\hbar/l_B},
\end{eqnarray}
which satisfy the commutation relation $[a,a^\dagger] = 1$. Using these operators, (the kinetic part of) the Hamiltonian becomes $H=\hbar \omega_c \sum_i \big( a_i^\dagger a_i + \frac{1}{2} \big)$, where $\omega_c=\frac{eB}{mc}$ is the cyclotron frequency. Eigen subspaces form the Landau levels (LLs), which are determined by the dynamical momenta degrees of freedom only. Each LL has extensive degeneracy that can be distinguished by the consideration of the guiding-center coordinates,
\begin{equation}
\R = (X,Y) =  \Big( x+ \frac{l_B^2}{\hbar} \pi_y, y- \frac{l_B^2}{\hbar} \pi_x \Big).
\end{equation} 
The guiding-center coordinates are chosen in such a way that they are linear combinations of coordinate operators and the dynamical momenta, while having trivial commutation relations with the dynamical momenta. The guiding-center coordinates have the commutation relation $[X,Y] = -il_B^2$. It is convenient to introduce operators
\begin{eqnarray}
b &=& \frac{1}{\sqrt{2}} \frac{X-iY}{l_B} \nonumber \\
b^\dagger &=& \frac{1}{\sqrt{2}} \frac{X+iY}{l_B},
\end{eqnarray}
which have the commutation relation $[b,b^\dagger] = 1$. We \textit{define} an angular momentum operator by
\begin{equation}
L_z = \hbar \big( b^\dagger b - a^\dagger a\big)
\label{angular_momentum}
\end{equation}
which commutes with (the kinetic part of) the Hamiltonian. The above definition of angular momentum is different from the usual definition of the angular momentum $x p_y - y p_x$~\cite{FootnoteAng}. However, this definition allows us to express the physical operators, such as relative angular momentum operators, using the second quantized operators only without referring to a specific gauge. Together with $a$ and $b$, we can construct the complete basis 
\begin{equation}
\ket{n,m} = \frac{(b^\dagger)^{m+n}}{\sqrt{(m+n)!}} \frac{(a^\dagger)^n}{\sqrt{n!}} \ket{0,0},
\end{equation}
where $\ket{0,0}$ is the vacuum state annihilated by $a$ and $b$, $n\in \{0,1,2,\dots\}$ is the Landau-level index, and $m\in\{-n,-n+1,\dots,0,1,\dots\}$ is the angular momentum index. Without interaction, the Landau-level index $n$ determines the eigenenergy $\big(n+\frac{1}{2}\big) \hbar \omega_c$, and the energy gap $\hbar \omega_c$ is proportional to $B$. 

In the limit of strong magnetic field, electrons in the FQH sit only in the lowest Landau level (LLL) and the guiding-center coordinates are the only relevant physical degrees of freedom.  These degrees of freedom provide a faithful representation to describe interactions in the LLL. We share the same point of view as previous literature~\cite{murthy2003review, haldane2011geometrical, haldane2011selfdual, seidel2013pairing, seidel2015algebraic, seidel2015bosonization} in that the guiding-center degrees of freedom are the necessary and sufficient degrees of freedom to describe electron states in the (idealized) FQH.  (Essentially we do not want to go back to the full electron Hilbert space which requires additionally specification of the dynamical momenta degrees of freedom~\cite{seidel2013pairing, seidel2015algebraic, seidel2015bosonization}.)

The interaction Hamiltonian should be expressed in terms of second quantized (guiding-center) operators. Haldane's pseudopotentials~\cite{haldane1983sphere}, which are relative angular momentum projectors, span the (sub)space of interaction Hamiltonians. The notion of pseudpotential makes sense without ever specifying the gauge of the magnetic field since it is built upon the relative angular momentum.  As an exception, the symmetric gauge will be used to find coordinate representations~\cite{lee2013pseudopotential} of pseudopotentials, as it requires specification of the action of the interaction Hamiltonian in the full (position) electron Hilbert space.

\subsection{Orbitals in various gauges and geometries}
Second-quantized expressions are already sufficient to capture the physics of the FQHE, but it is often more intuitive to use first-quantized expressions. The first step towards using the first-quantized language is to specify the gauge.

We start with the most familiar one, the symmetric gauge $\bf{A} = B\big(\frac{y}{2}, -\frac{x}{2}\big)$. The complete eigenstates are given by~\cite{jainbook}
\begin{equation}
\ip{\r}{n,m} = \frac{(-1)^n}{\sqrt{2\pi l_B^2}}\sqrt{\frac{n!}{2^m (m+n)!}} z^m L_n^m \Big( \frac{|z|^2}{2} \Big) e^{-\frac{1}{4}|z|^2}, \nonumber
\end{equation}
where we have defined a dimensionless complex coordinate $z=(x+iy)/l_B$, $L_n^m$ is the associated Laguerre polynomial, and $n=0$ corresponds to a familiar LLL orbitals  $\phi_m(\r) = \frac{1}{\sqrt{2\pi 2^m m! l_B^2}} z^m e^{-\frac{1}{4}|z|^2}$. In this gauge, the usual rotational symmetry becomes the symmetry of the Hamiltonian and the angular momentum Eq.~(\ref{angular_momentum}) becomes identical to the angular momentum in the usual sense, $L_z = \hbar \big( b^\dagger b - a^\dagger a \big) = x p_y - y p_x$.

The second gauge is the Landau gauge $\bf{A} = B(0,-x)$. This gauge is useful when considering a cylinder obtained by compactifying a plane in the $y$-direction. After the gauge transformation from the symmetric gauge to the Landau gauge, a LLL basis can be written as
\begin{equation}
\phi_m^\textrm{Landau} (\r) = \frac{1}{\sqrt{2\pi 2^m m! l_B^2}} \bigg( \frac{x+iy}{l_B} \bigg)^n e^{-\frac{1}{4} \frac{x^2+y^2}{l_B^2} + \frac{i}{2} \frac{xy}{l_B^2}},
\label{basis_angular_Landau}
\end{equation}
where $m$ is an angular momentum eigenvalue of Eq.~(\ref{angular_momentum}). (We often drop the superscript ``Landau'' whenever the gauge choice is clear.) In addition, since the Landau gauge has translational symmetry in $y$-direction, it is useful to consider eigenstates of LLL given by
\begin{equation}
\ip{\r}{\psi_k} = \psi_k(\r) = \frac{1}{\pi^{1/4} l_B} e^{iky} e^{-\frac{(x-k l_B^2)^2}{2 l_B^2}},
\label{basis_Landau}
\end{equation}
where $k l_B \in \mathbbm{R}$. This basis satisfy the orthonormality condition $\int d^2 \r \psi_{k'}^*(\r) \psi_k (\r) = 2\pi \delta (l_B (k' - k) )$ and form a completeness basis of the LLL $\int \frac{d(l_B k)}{2\pi} \op{\psi_k}{\psi_k} = \openone |_{\mathcal{H}_\textrm{LLL}^\textrm{Landau}}$. The basis transformation matrix element between Eq.~(\ref{basis_angular_Landau}) and Eq.~(\ref{basis_Landau}) is given by
\begin{equation}
\ip{\psi_k}{\phi_m} = \sqrt{\frac{\pi^{1/2}}{2^{m-1}m!}} H_m \big( l_B k \big) e^{-\frac{1}{2} l_B^2 k^2}.
\label{basis_change_Landau}
\end{equation}

When compactifying $y$-direction of a plane via $y \sim y+L_y$, we get a cylinder together with the basis
\beqn
\ip{\r}{\psi_n} = \psi_n(\r) = \frac{\sqrt{\kappa}}{\sqrt{2\pi^{3/2} l_B^2}} e^{i\kappa n \frac{y}{l_B}} e^{-\frac{(x-\kappa n l_B)^2}{2 l_B^2}}, \nonumber
\label{basis_cylinder}
\eeqn
where $\kappa = \frac{2\pi l_B}{L_y}$ is the dimensionless inverse radius of the cylinder and $n\in \mathbbm{Z}$. We have the orthonormality condition $\int_\textrm{cyl} d^2 \r \psi_{n_1}^*(\r) \psi_{n_2} (\r) = \delta_{n_1,n_2}$ and the completeness relation $\sum_{n=-\infty}^{\infty} \op{\psi_n}{\psi_n} = \openone |_{\mathcal{H}_{\textrm{LLL}}^{\textrm{cyl}}}$.

The torus geometry requires further compactification in $x$-direction via $x \sim x+L_x$. A basis is given by the linear superposition of the orbitals of the cylinder: $\sum_{s\in \mathbbm{Z}} \psi_{n+s N_\phi}$, where $N_\phi = \frac{L_x L_y}{2 \pi l_B^2} \in \mathbbm{Z}$ is the total flux through torus and equals the total number of orbitals.

Unlike other geometries, the spherical geometry has full rotational symmetry generated by $L_+, L_-, L_z$, which is defined in Eq.~(\ref{sphere_angular_operators}). When a magnetic monopole with magnetic charge $2S$ is placed in the origin, the LLL is spanned by normalized orbitals $\phi_M (\theta,\phi) = \sqrt{\frac{2S+1}{4\pi}} [u ]^S_M$~\cite{haldane1983sphere, jainbook, haxton2016}, where $M \in \{-S,-S+1,\dots,S\}$ and 
\begin{equation}
[u ]^S_M = \sqrt{\frac{(2S)!}{(S+M)! (S-M)!}} (u_{1/2})^{S+M} (u_{-1/2})^{S-M}
\end{equation}
with
\begin{eqnarray}
u_m(\theta,\phi) = 
\left\{
\begin{array}{lc}
\cos(\theta/2) e^{i\phi/2}, & m=\frac{1}{2} \\
\\
\sin(\theta/2) e^{-i\phi/2}, & m=-\frac{1}{2}
\end{array} \right. . \nonumber
\end{eqnarray}
$\{ [u]^S_M \}_{M=-S,\dots,S}$ forms a $SU(2)$ spin $S$ representation under the generators $L_\pm, L_z$, where $\vec{L} = \vec{r} \times \vec{\pi} + \hbar S \hat{\Omega}$ and $\hat{\Omega}$ is the unit normal vector on the sphere. We often identify $N_\phi \equiv 2S$ as the total flux of the sphere. A stereographic projection from sphere to plane amounts to mapping from $\phi_M(\theta,\phi)$ to $\phi_m(z)$ via $M = m - S$ up to normalization factors. 

\section{Pseudopotentials}
\label{appendix_pseudopotentials}
We now systematically construct and classify the Haldane's pseudopotentials~\cite{haldane1983sphere} in terms of the \textit{second-quantization}, which gives a faithful representation using the guiding-center degrees of freedom only. We defer a more familiar representations involving the position and momentum coordinates to section~(\ref{pseudopotentials_coordinates}). 

While our construction essentially gives identical results to the previous literature on pseudopotentials~\cite{simon2007pseudopotentials, lee2013pseudopotential, lee2015geometric}, we strictly follow the (canonical) construction of pseudopotentials in Ref.~\onlinecite{simon2007pseudopotentials} in which the first quantized orbitals in the symmetric gauge are used to construct pseudopotentials. Our construction can be viewed as the second-quantized analog of Ref.~\onlinecite{simon2007pseudopotentials}, so that we explicitly show the one-to-one correspondence of pseudopotentials in various geometries.

\subsection{Jacobi coordinate system}
We are interested in classifying the translationally invariant many-body interactions which depend on the relative motion degrees of freedom only. We employ the Jacobi coordinate system to systematically express many-body interactions. In order to motivate the Jacobi transformation, we momentarily go back to coordinate representations by restoring the dynamical momenta degrees of freedom. One can skip the following discussion and jump directly to Eq.~(\ref{Jacobi_operators}), if one only cares about the guiding-center degrees of freedom.

The Jacobi coordinates associated with $N$ particles with coordinates $\r_1,\dots,\r_N$ is given by
\begin{eqnarray}
\left\{
\begin{array}{l}
\R_{\textrm{CM}} = \frac{\r_1+\dots+\r_N}{N} \\
\u_1 = \r_1-\r_2 \\
\u_2 = \frac{\r_1+\r_2 -2\r_3}{2} \\
\vdots \\
\u_{N-1} = \frac{\r_1+\dots+\r_{N-1}-(N-1)\r_N}{N-1},
\end{array} \right.
\label{Jacobi_coordinates}
\end{eqnarray}
which consist of the center-of-mass (CM) coordinate and the relative coordinates. Let's denote the coordinate transformation matrix between the ordinary coordinates and the Jacobi coordinates by $\bf{M}$, i.e., $\u_i = \sum_j [\bf{M}]_{ij} \r_j$. (We may identify $\u_N \equiv \R_\textrm{CM}$.) For sake of simplicity, we momentarily assume the gauge is given by the symmetric gauge. Using the coordinate transformation and $\frac{\partial}{\partial \r_i} = \sum_j [\bf{M}^T]_{ij} \frac{\partial}{\partial \u_j}$, the kinetic part of the Hamiltonian becomes~\cite{jainbook,lee2013pseudopotential}
\begin{eqnarray}
H_\textrm{kin} &=& \frac{\hbar \omega_c}{2} \sum_{i=1}^N \bigg[ \bigg( \frac{l_B}{i} \frac{\partial}{\partial x_i} + \frac{y_i}{2 l_B} \bigg)^2 + \bigg( \frac{l_B}{i} \frac{\partial}{\partial y_i} - \frac{x_i}{2 l_B} \bigg)^2 \bigg] \nonumber \\
&=& \frac{\hbar \omega_c}{2} \sum_{i=1}^N \bigg[ \bigg( \frac{l_i}{i} \frac{\partial}{\partial (\u_i)_x} + \frac{(\u_i)_y}{2 l_i} \bigg)^2 \nonumber \\
& & \qquad \qquad \qquad \quad + \bigg( \frac{l_i}{i} \frac{\partial}{\partial (\u_i)_y} - \frac{(\u_i)_x}{2 l_i} \bigg)^2 \bigg], 
\end{eqnarray}
where we have introduced the CM and rel length scales $(l_\textrm{CM}; l_i) = (\sqrt{\lambda_\textrm{CM}} l_B; \sqrt{\lambda_1} l_B, \dots, \sqrt{\lambda_{N-1}} l_B)$ with $\lambda_\textrm{CM} = \frac{1}{N}$ and $\lambda_n = \frac{n+1}{n}$. Note that the CM length scale \textit{depends} on the number of particles $N$, while rel length scales are \textit{independent} of $N$. We broadly follow the notations in Ref.~\onlinecite{lee2013pseudopotential} but differences occur due to our explicit coordinate choice. In the Jacobi coordinates, the normalized LLL orbitals are given by
\begin{align}
\phi_M^\textrm{CM} (\R_N) &= \frac{\big(Z_N/\sqrt{\lambda_\textrm{CM}}\big)^M}{\sqrt{2\pi 2^M M! l_\textrm{CM}^2}} e^{-\frac{1}{4} |\R_N|^2/l_\textrm{CM}^2} \\
\phi_m^\textrm{rel} (\u_i) &= \frac{\big( z_i/\sqrt{\lambda_i} \big)^m}{\sqrt{2\pi 2^m m! l_i^2}} e^{-\frac{1}{4} |\u_i|^2/l_i^2} ,
\end{align}
where $Z_N = \big( (\R_N)_x + i (\R_N)_y \big)/l_B$ and $z_i = \big( (\u_i)_x + i (\u_i)_y)/l_B$.

Moreover, $a$ operators are transformed as
\begin{eqnarray}
a^\textrm{CM} &=& \frac{1}{\sqrt{\lambda_\textrm{CM}} } \big( \frac{1}{N} a_1 + \dots + \frac{1}{N} a_N \big) \nonumber \\
a^{\textrm{rel}}_i &=& \frac{1}{\sqrt{\lambda_i}} \sum_{j=1}^N [\bf{M}]_{ij} a_j,
\end{eqnarray}
and the same formulas hold for $b$:
\begin{eqnarray}
b^\textrm{CM} &=& \frac{1}{\sqrt{\lambda_\textrm{CM}} } \big( \frac{1}{N} b_1 + \dots + \frac{1}{N} b_N \big) \nonumber \\
b^{\textrm{rel}}_i &=& \frac{1}{\sqrt{\lambda_i}} \sum_{j=1}^N [\bf{M}]_{ij} b_j.
\label{Jacobi_operators}
\end{eqnarray}
All operators satisfy the canonical commutation relation $[a,a^\dagger] = [b,b^\dagger]=1$. Let's now forget about the Jacobi coordinates, and \textit{define} the Jacobi transformation of second quantized operators as Eq.~(\ref{Jacobi_operators}) from the onset. Length scales $(l_\textrm{CM};l_i)$ are chosen in such a way that the operators satisfy the canonical commutation relation. This \textit{definition} doesn't require any restrictions on the gauge choice. We use this transformed Jacobi guiding-center degrees of freedom to construct many-body pseudopotentials.

\subsection{Clebsch-Gordan coefficients}
There exist two basis for the Hilbert space of $N$ electrons - one is the ordinary orbital basis, $\Big\{ \ket{m_1,\dots,m_N} = \frac{(b_1^\dagger)^{m_1}}{\sqrt{m_1!}} \cdots \frac{(b_N^\dagger)^{m_N}}{\sqrt{m_N!}} \ket{0,\dots,0}\Big\}$ and the other is the Jacobi transformed basis, $\Big\{ \ket{M,m'_1,\dots,m'_{N-1}} = \frac{(b^{\textrm{CM} \dagger})^{M}}{\sqrt{M!}} \frac{(b^{\textrm{rel} \dagger}_{1})^{m'_{1}}}{\sqrt{m'_{1}!}} \cdots \frac{(b^{\textrm{rel} \dagger}_{N-1})^{m'_{N-1}}}{\sqrt{m'_{N-1}!}} \ket{0,\dots,0} \Big\}$ with $m_i,M,m'_i \in \{0,1,2,\dots\}$. The basis change matrix, which we call the Clebsch-Gordan coefficients, is given by $\ip{M,m'_1,\dots,m'_{N-1}}{m_1,\dots,m_N}$. The nomenclature becomes clear once we compare the Clebsch-Gordan coefficient of the plane and that of the sphere, which is done in Eq.~(\ref{clebsch_gordan_sphere_plane}). Using Eq.~(\ref{Jacobi_operators}), the two-body Clebsch-Gordan coefficient is given by
\begin{widetext}
\begin{eqnarray}
\ip{M,m}{m_1,m_2} &=& \bra{0,0} \frac{1}{\sqrt{M!}} \Big( \frac{b_1 + b_2}{\sqrt{2}} \Big)^M \frac{1}{\sqrt{m!}} \Big( \frac{b_1 - b_2}{\sqrt{2}} \Big)^m \ket{m_1,m_2} \nonumber \\
&=& \delta_{M+m,m_1+m_2} \sqrt{\frac{m_1! m_2!}{2^{M+m} M! m!}} \sum_{K=0}^M \sum_{k=0}^m (-1)^k \binom{M}{K} \binom{m}{k} \delta_{K+k,m_2} \nonumber \\
&=& \delta_{M+m,m_1+m_2} \sqrt{\frac{m_1! m_2!}{2^{M+m} M! m!}} \frac{M!}{m_2!} \frac{{}_2 F_1 (-m,-m_2;M-m_2+1,-1)}{\Gamma(M-m_2+1)} \nonumber \\
&=& \delta_{M+m,m_1+m_2} \sqrt{\frac{M! m!}{2^{M+m} m_1! m_2!}} P_m^{(m_1-m,-m_1-m_2-1)} (3),
\end{eqnarray}
where $\delta$ is the Kronecker delta, ${}_2F_1$ is the hypergeometric function having $\Gamma$ function in the denominator as a regulator, and $P_n^{(\alpha,\beta)}$ is the Jacobi polynomial. After similar computations, the three-body Clebsch-Gordan coeficient is given by
\begin{eqnarray}
\ip{M,m'_1,m'_2}{m_1,m_2,m_3} &=& \delta_{M+m'_1+m'_2,m_1+m_2+m_3} \sqrt{\frac{m_1! m_2! m_3!}{2^{m'_1+m'_2} 3^{M+m'_2} M! m'_1! m'_2!}} \frac{M!}{m_3!} \frac{{}_2F_1 (-m'_2,-m_3;M-m_3+1,-2)}{\Gamma(M-m_3+1)} \nonumber \\
& & \qquad \qquad \qquad \times \frac{(M+m'_2-m_3)!}{m_2!} \frac{{}_2F_1 (-m_2,-m'_1;M+m'_2-m_2-m_3+1,-1)}{\Gamma(M+m'_2-m_2-m_3+1)} \nonumber \\
&=& \delta_{M+m'_1+m'_2,m_1+m_2+m_3} \sqrt{\frac{m_1! m_2! m_3!}{2^{m'_1+m'_2} 3^{M+m'_2} M! m'_1! m'_2!}} \frac{M! m'_2!}{m_3! (m_1+m_2-m'_1)!} \nonumber \\
& & \qquad \qquad \qquad \qquad \qquad \qquad \times P_{m'_2}^{(M-m_3,-M-m'_2-1)} (5) \, P_{m_2}^{(m_1-m'_1,-m_1-m_2-1)} (3).
\end{eqnarray}
\end{widetext}

\subsection{Relative angular momentum eigenstates}
Haldane's pseudopotentials~\cite{haldane1983sphere, simon2007pseudopotentials} are identified as the projections onto the relative angular momentum eigenstates. In the following, we give a precise definition of the relative angular momentum eigenstates which in turn are the eigenstates of the relative angular momentum operator. From now on, we set $\hbar=1$.

The total angular momentum operator for $N$ electrons is $L_z^\textrm{tot} = \sum_{i=1}^N b_{i}^\dagger b_{i}$. After the Jacobi transformation Eq.~(\ref{Jacobi_operators}), the total angular momentum operator becomes $L_z^\textrm{tot} = (b^{\textrm{CM}})^\dagger b^{\textrm{CM}} + \sum_{i=1}^{N-1} (b_{i}^\textrm{rel})^\dagger b_{i}^\textrm{rel}$. It is then natural to \textit{define} the relative angular momentum operator of $N$-particle by
\begin{equation}
L_z^\textrm{rel} = \sum_{i=1}^{N-1} (b_{i}^\textrm{rel})^\dagger b_{i}^\textrm{rel},
\end{equation}
and the CM angular momentum operator of $N$-particle by $L_z^\textrm{CM} = (b^{\textrm{CM}})^\dagger b^{\textrm{CM}}$. The eigenstates of the relative angular momentum operators are precisely the Jacobi transformed basis states $\ket{M,m_1',\dots,m_{N-1}'}$ with the relative angular momentum $m'_1 + \dots +m'_{N-1}$. (The CM angular momentum is $M$ in this case.) However, complications arise when considering the \textit{bosonic} (or \textit{fermionic}) nature of identical electrons. This restricts the basis to be symmetric (antisymmetric) under the exchange of the bosonic (fermionic) electrons. From now on, we consider the bosonic case only, while the fermionic case can be considered analogously.

Let's work in the the CM angular momentum $0$ sector and find the relative angular momentum eigenstates in this sector. Upon applying $\frac{1}{\sqrt{M!}} (b^{\textrm{CM}\dagger})^M$, we get the corresponding relative angular momentum eigenstate with CM angular momentum $M$. Since we already know unsymmetrized relative angular momentum eigenstates, the symmetrized eigenstates follow from symmetrization. For example, from a relative angular momentum eigenstate $\ket{0,m_1',\dots,m_{N-1}'}$ which has a relative angular momentum $m=m_1' + \dots + m_{N-1}'$, the symmetrized eigenstate is given by
\begin{align}
\mathcal{S} \Big[ \ket{0,&m_1',\dots,m_{N-1}'} \Big] \nonumber \\
\propto &\mathcal{S} \Big[ ( b_1^\dagger - b_2^\dagger )^{m_1'} (b_1^\dagger + b_2^\dagger - 2 b_3^\dagger)^{m_2'} \dots \Big] \ket{0,\dots,0}, \nonumber
\end{align}
where $\mathcal{S}$ symmetrizes the particle indices in $b^\dagger$. So the relative angular momentum eigenstates are obtained by acting homogeneous symmetric polynomials of creation \textit{operators} $\{ b^\dagger_i \}$ on the vacuum state. Since we are working in the $M=0$ sector, the homogeneous symmetric polynomials creating relative angular momentum eigenstates are precisely the \textit{translationally invariant} homogeneous symmetric polynomials, which are classified in Ref.~\onlinecite{simon2007pseudopotentials}.  Ref.~\onlinecite{simon2007pseudopotentials} uses the advantage of the polynomial structure of the orbital basis in the coordinate representation using the symmetric gauge, but in our work, the same structure arises at the level of second quantized operators without ever specifying the gauge. Let's denote an orthonormalized relative angular momentum eigenstate by $\ket{m,a}_\textrm{rel}$ (which depends on $N$, the number of particles forming the eigenstate and we often drop the subscript ``rel''), where $a \in \{1,\dots,D_{m,N}\}$ accounts for the degeneracy. Using the prescription in Ref.~\onlinecite{simon2007pseudopotentials}, we list the first few relative angular momentum eigenstates in TABLE~\ref{Table_rel_ang}.

The second quantized operators $b_j$ and $b_j^\dagger$ used so far are associated with the \textit{orbital} basis of particle $j$. When dealing with the system of identical particles, it is convenient to use the creation and annihilation operators $c_m$ and $c_m^\dagger$ associated with \textit{occupation number} basis, where $m\in\{0,1,2,\dots\}$ and the operators satisfy the commutation relation $[c_m,c_{m'}^\dagger] = \delta_{m,m'}$. For example, the total angular momentum operator can be written as: $L_z^\textrm{tot} = \sum_{i=1}^N b_i^\dagger b_i \to \sum_{m=0}^\infty m c_m^\dagger c_m$.

\begin{widetext}
\onecolumngrid
\begin{table}[t]
\begin{tabular}{c|c|c}
\hline \hline
\multirow{2}{*}{Eigenvalue} & \multicolumn{2}{c}{Eigenstates} \\ \cline{2-3}
\multicolumn{1}{c|}{} & \multicolumn{1}{c|}{$\ket{m_1'}$} & \multicolumn{1}{c}{$\ket{m_1,m_2}$} \\
\hline
$m$ & $\ket{m}$ & \(\displaystyle{ \frac{1}{\sqrt{2^m}} \sum_{k=0}^m \frac{(-1)^k \sqrt{m!}}{\sqrt{k! (m-k)!}} \ket{m-k,k} }\) \\
\hline \hline
\end{tabular}
\\
(a) $N=2$ \\
\medskip
\begin{tabular}{c|c|c}
\hline \hline
\multirow{2}{*}{Eigenvalue} & \multicolumn{2}{c}{Eigenstates} \\ \cline{2-3}
\multicolumn{1}{c|}{} & \multicolumn{1}{c|}{$\ket{m'_1,m'_2}$} & \multicolumn{1}{c}{$\ket{m_1,m_2,m_3}$} \\
\hline
$0$ & $\ket{0,0}$ & $\ket{0,0,0}$ \\
\hline
$2$ & $\frac{1}{\sqrt{2}} \ket{2,0}+ \frac{1}{\sqrt{2}} \ket{0,2}$ & $\sqrt{\frac{2}{3}} \frac{\ket{2,0,0} + \ket{0,2,0} + \ket{0,0,2}}{\sqrt{3}} - \sqrt{\frac{1}{3}} \frac{\ket{1,1,0} + \ket{1,0,1} + \ket{0,1,1}}{\sqrt{3}} $ \\
\hline
$3$ & $\frac{\sqrt{3}}{2} \ket{2,1} - \frac{1}{2} \ket{0,3}$ & $\frac{\sqrt{2}}{3} \frac{\ket{3,0,0}+\ket{0,3,0}+\ket{0,0,3}}{\sqrt{3}} -\frac{1}{\sqrt{3}} \frac{\ket{2,1,0}+\ket{2,0,1}+\ket{0,2,1}+\ket{1,2,0}+\ket{1,0,2}+\ket{0,1,2}}{\sqrt{6}} + \frac{2}{3} \ket{1,1,1}$ \\
\hline \hline
\end{tabular} \\
(b) $N=3$ \\
\caption{Relative angular momentum eigenenstates in the Jacobi and the product basis in the CM angular momentum $0$ sector of two and three particles. We suppress the CM angular momentum index in the Jacobi basis. Only even $m$ gives a nonvanishing state when $N=2$.}
\label{Table_rel_ang}
\end{table}
\end{widetext}

\subsection{Pseudopotentials in the second quantization}
Having defined the relative angular momentum eigenstates, one can construct associated projection operators, so-called the Haldane pseudopotentials~\cite{haldane1983sphere, simon2007pseudopotentials}. The pseudopotentials form complete basis in the sense that any symmetric N-body interaction $V_N$, which commutes with the CM angular momentum operator and the rel angular momentum operator (so the angular momenta remain good quantum numbers) and acts trivially in the CM sector, can be written as sums of pseudopotentials:
\begin{equation}
V_N = \sum_{m=0}^\infty \sum_{a,b=1}^{D_{m,N}} \oev{m,a}{V_N}{m,b} \openone_{\mathcal{H}_{\textrm{LLL}}^{\textrm{CM}}} \otimes \op{m,a}{m,b},
\end{equation}
where $\openone_{\mathcal{H}_{\textrm{LLL}}^{\textrm{CM}}} = \sum_{M=0}^\infty \op{M}{M}$ is the identity operator in the CM sector. The $N$-body pseudopotential is given by
\begin{equation}
P_N^{(m,a)} = \frac{1}{N!} \sum_{M=0}^\infty T_M^\dagger T_M,
\label{pseudopotential_projection}
\end{equation}
where we have defined $T_M = \big[T_N^{(m,a)} \big]_M = \sum_{m_i =0}^\infty \ip{M;(m,a)}{m_1,\dots,m_N} c_{m_1} \cdots c_{m_N}$ and $a \in \{1,\dots,D_{m,N}\}$ accounts for the degeneracy in the $N$-body relative angular momentum $m$ subspace. 

A systematic procedure for constructing the pseudopotentials can be summarized as follows: (a) Find the rel angular momentum eigenstate $\ket{m;a}$ in the CM angular momentum $0$ sector. (b) Find the corresponding rel angular momentum eigenstate in the CM angular momentum $M$ sector by applying $\frac{1}{\sqrt{M!}} \big( \frac{b_1^\dagger + \dots + b_N^\dagger}{\sqrt{N}} \big)^M$ to $\ket{m,a}$ or using the Clebsch-Gordan coefficients to construct $T_M$. (c) The pseudopotential follows from Eq.~(\ref{pseudopotential_projection}). 

For the sake of completeness, we also present the occupation number operator based construction of the pseudopotentials. The idea is to represent $b^\textrm{CM} = \frac{b_1 + \dots + b_N}{\sqrt{N}}$ in terms of the occupation number operators. The $N$-body CM angular momentum operator can be written as 
\begin{eqnarray}
L_z^\textrm{CM} &=& \frac{1}{N} L_{+}^\textrm{CM} L_{-}^\textrm{CM} \nonumber \\
L_{+}^\textrm{CM} &=& \sum_{m=0}^\infty \sqrt{m+1} c_{m+1}^\dagger c_m \nonumber \\
L_{-}^\textrm{CM} &=& \sum_{m=0}^\infty \sqrt{m+1} c_m^\dagger c_{m+1},
\end{eqnarray}
where the operators $L_z^\textrm{CM}$ and $L_{\pm}^\textrm{CM}$ do \textit{not} satisfy the $SU(2)$ algebra. (We could construct operators satisfying the $SU(2)$ algebra in analogy with the sphere at the expense of introducing a finite cut-off in the orbitals.) Starting from $T_0 = \big[T_N^{(m,a)} \big]_0$, which is expressed in terms of occupation number operators, the following recursion relation holds:
\begin{equation}
T_{M+1}^\dagger = \frac{1}{\sqrt{N}} \Big[ L_{+}^\textrm{CM} , T_M^\dagger \Big].
\end{equation}
The (normalized) angular momentum eigenstate is given by $\ket{M, (m,a)} = \frac{1}{\sqrt{N!}}T_M^\dagger \ket{0}$.

Suppose we want to find a $zero$ energy ground state wavefunction $\ket{\Psi_{N_e}}$ (where $N_e$ is the number of electrons) of a Hamiltonian $H= \sum_{\alpha=1}^l P_{N_\alpha}^{(m_\alpha,a_\alpha)}$. As the Hamiltonian is written in terms of sums of positive projectors, a zero energy state would be annihilated by individual projectors in the Hamiltonian. If we further assume to have maximum possible orbital, say $N_\phi+1$, finding the ground state amounts to solving a system of homogeneous $linear$ equations~\cite{chandran2011bulkedge}: $\big[ T_{N_\alpha}^{(m_\alpha,a_\alpha)} \big]_M \ket{\Psi_{N_e}} = 0$, $\forall \alpha \in \{1,2,\dots,l\}$, $\forall M \in \{0,1,2,\dots, N_\phi-m_\alpha \}$. This is a great simplification compared to numerically expensive exact diagonalization.

\subsection{Pseudopotentials in the cylinder geometry}
In this section, we derive the second quantized Hamiltonian in the cylinder geometry. For this purpose, we rewrite the pseudopotentials in Eq.~(\ref{pseudopotential_projection}) in terms of the basis in Eq.~(\ref{basis_Landau}). Expressions for the cylinder geometry follows when compactifying the plane properly. During the computation of the matrix element of the Hamiltonian, we encounter
\begin{widetext}
\begin{align}
& \sum_{M=0}^\infty \ip{k_1',\dots,k_N'}{M,m'_1,\dots,m'_{N-1}} \ip{M,m_1,\dots,m_{N-1}}{k_1,\dots,k_N} \nonumber \\
& \quad = \bigg(\prod_{i=1}^N \int d^2 \r'_i \psi_{k'_i}^* (\r'_i) \int d^2 \r_i \psi_{k_i} (\r_i) \bigg) \bigg(\sum_{M=0}^\infty \phi_M^\textrm{CM} (\mathbf{R}') \big(\phi_M^{\textrm{CM}} (\mathbf{R})\big)^*\bigg) \bigg( \prod_{j=1}^{N-1} \phi_{m'_j}^{\textrm{rel},j} (\u'_j) \big(\phi_{m_j}^{\textrm{rel},j} (\u_j)\big)^* \bigg) \nonumber \\
& \quad = \bigg(\prod_{i=1}^N \int d^2 \r'_i \psi_{k'_i}^* (\r'_i) \int d^2 \r_i \psi_{k_i} (\r_i) \bigg) \bigg( \frac{N}{2\pi l_B^2} e^{\frac{N (Z^* Z' - i (Z)_x (Z)_y + i (Z')_x (Z')_y )}{2} - \frac{N}{4} (|Z|^2+|Z'|^2)} \bigg) \bigg( \prod_{j=1}^{N-1} \phi_{m'_j}^{\textrm{rel},j} (\u'_j) \big(\phi_{m_j}^{\textrm{rel},j} (\u_j)\big)^* \bigg) \nonumber \\
& \quad = \sqrt{N} e^{l_B^2 \big(\frac{k_1+\dots+k_N}{\sqrt{N}}\big)^2} e^{-\l_B^2 (k_1^2+\dots+k_N^2)} 2\pi \delta \big( l_B(k_1+\dots+k_N) - l_B(k'_1+\dots+k'_N) \big) \nonumber \\
& \qquad \times \prod_{i=1}^{N-1} \bigg( \frac{2 \pi^{1/2}}{\sqrt{2^{m_i+m'_i} m_i! m'_i!}} H_{m_i} \Big(\frac{l_B \big(k_1+\dots+k_i - i k_{i+1} \big)}{\sqrt{i(i+1)}}  \Big) H_{m'_i} \Big(\frac{l_B \big(k'_1+\dots+k'_i - i k'_{i+1} \big)}{\sqrt{i(i+1)}} \Big) \bigg),
\label{matrix_element_Landau}
\end{align}
where $\{\R,\u\}$ and $\{\r\}$ are related by the Jacobi coordinate transformation Eq.~(\ref{Jacobi_coordinates}), $Z=\frac{(\R)_x+i(\R)_y}{l_B}$, $(Z)_x = (\R)_x/l_B$, and so on, and $H_m$ is the $m$-th Hermite polynomial. In Eq.~(\ref{matrix_element_Landau}), $\ket{M,m_1,\dots,m_{N-1}}$ is the Jacobi transformed basis and $\phi$ and $\psi$ are eigenstates in \textit{Landau} gauge, i.e., (Jacobi transformed) Eq.~(\ref{basis_angular_Landau}) and Eq.~(\ref{basis_Landau}). 

We again emphasize that the pseudopotential in Eq.~(\ref{pseudopotential_projection}) is independent of the gauge choice. Whereas, we used the Landau gauge to calculate matrix element in Eq.~(\ref{matrix_element_Landau}). Alternatively, one can simply think of Eq.~(\ref{matrix_element_Landau}) as an abstract unitary basis transformation from the relative angular momentum basis to another basis guided by the unitary transformation given in Eq.~(\ref{basis_change_Landau}). This particular unitary transformation gives expressions for the pseudopotentials which can be interpreted as interaction Hamiltonians in the cylinder and the torus geometry, i.e., respect desired symmetries.

In the following, we present explicit expressions for the two-body and the first few three-body pseudopotentials in a cylinder. The cylinder is obtained by compactifying a plane in $y$-direction via $y \sim y+L_y$. Further pseudopotentials can be systematically derived using Eq.~(\ref{matrix_element_Landau}) starting from the pseudopotentials in the plane geometry. The two-body pseudopotential is given by
\begin{equation}
P_2^m = \frac{1}{2!} \sum_{R \in \mathbbm{Z}/2} T_R^\dagger T_R,
\end{equation}
where 
\begin{equation}
T_R = \big[ T_2^{m} \big]_R = \Big(\frac{2}{\pi}\Big)^{1/4} \sqrt{\kappa} \sum_{\substack{-\infty <r < \infty \\r+R \in \mathbbm{Z}}} \frac{1}{\sqrt{2^m m!}} H_m \big( \kappa \sqrt{2} r \big) e^{-\kappa^2 r^2} c_{R+r} c_{R-r}.
\end{equation}

We present the three-body pseudopotentials for $m=0,2,3$. This can be derived using Eq.~(\ref{matrix_element_Landau}) and Table~\ref{Table_rel_ang} (b). The three-body pseudopotential, $P_3^m$ ($m=0,2,3$), in the cylinder geometry is
\begin{equation}
P_3^m = \frac{1}{3!} \sum_{R \in \mathbbm{Z}/3} T_R^\dagger T_R,
\end{equation}
where $T_R = \big[T_3^m \big]_R$ and
\begin{align}
\big[T_3^0 \big]_R &= \Big(\frac{3}{\pi^2}\Big)^{1/4} \kappa \sum_{\substack{-\infty < r_1,r_2 <\infty \\ r_1+R, r_2+R \in \mathbbm{Z}}} e^{- \kappa^2 \big(r_1^2 + r_2^2 + r_1 r_2 \big)} c_{R+r_1} c_{R+r_2} c_{R -r_1-r_2} \\
\big[T_3^2 \big]_R &= \Big(\frac{3}{\pi^2}\Big)^{1/4} \kappa \sum_{\substack{-\infty < r_1,r_2 <\infty \\ r_1+R, r_2+R \in \mathbbm{Z}}}  \bigg( \frac{1}{4} H_2 \Big( \frac{\kappa (r_1-r_2)}{\sqrt{2}} \Big) + \frac{1}{4} H_2 \Big( \frac{\sqrt{3} \kappa (r_1+r_2)}{\sqrt{2}} \Big) \bigg) \nonumber \\
& \qquad \qquad \qquad \qquad \qquad \qquad \times e^{- \kappa^2 \big(r_1^2 + r_2^2 + r_1 r_2 \big)} c_{R+r_1} c_{R+r_2} c_{R -r_1-r_2} \\
\big[T_3^3 \big]_R &= \Big(\frac{3}{\pi^2}\Big)^{1/4} \kappa \sum_{\substack{-\infty < r_1,r_2 <\infty \\ r_1+R, r_2+R \in \mathbbm{Z}}} \bigg( \frac{\sqrt{3}}{8} H_2 \Big( \frac{\kappa (r_1-r_2)}{\sqrt{2}} \Big) H_1 \Big( \frac{\sqrt{3} \kappa (r_1+r_2)}{\sqrt{2}} \Big) - \frac{1}{8 \sqrt{3}}  H_3 \Big( \frac{\sqrt{3} \kappa (r_1+r_2)}{\sqrt{2}} \Big) \bigg) \nonumber \\
& \qquad \qquad \qquad \qquad \qquad \qquad \times e^{- \kappa^2 \big(r_1^2 + r_2^2 + r_1 r_2 \big)} c_{R+r_1} c_{R+r_2} c_{R -r_1-r_2}.
\end{align}
All pseudopotentials respect translational symmetry.

\subsection{Pseudopotentials in the torus geometry}
Pseudopotentials in the torus geometry follow directly from the cylinder geometry by compactifying $x$-direction via $x \sim x+L_x$. The two-body pseudopotential is given by
\begin{equation}
P_2^m = \frac{1}{2!} \sum_{2R = 1}^{2 N_\phi} T_R^\dagger T_R,
\end{equation}
where $m$ is the two-body relative angular momentum and 
\begin{equation}
T_R = \Big(\frac{2}{\pi}\Big)^{1/4} \sqrt{\kappa}	 \sum_{\substack{0 \le r < N_\phi \\ r+R \in \mathbbm{Z}}} \Bigg[ \sum_{s \in \mathbbm{Z}} \frac{1}{\sqrt{2^m m!}} H_m \big( \sqrt{2} \kappa (r + s N_\phi ) \big) e^{-\kappa^2 (r + s N_\phi)^2} \Bigg] c_{R+r} c_{R-r}.
\end{equation}
We have used the identification $c_{m+ N_\phi} = c_m$. The three-body pseudopotential is given by
\begin{equation}
P_3^{(m,a)} = \frac{1}{3!} \sum_{3R = 1}^{3N_\phi} T_R^\dagger T_R,
\end{equation}
where 
\begin{equation}
T_R = \Big(\frac{3}{\pi^2}\Big)^{1/4} \kappa \sum_{\substack{0 \le r_1,r_2 <N_\phi \\ r_1+R, r_2+R \in \mathbbm{Z}}} \Bigg[ \sum_{s_1,s_2 \in \mathbbm{Z}} t_3^{(m,a)} e^{-\kappa^2 \big( (r_1 + s_1 N_\phi)^2 + (r_2 + s_2 N_\phi)^2 + (r_1+ s_1 N_\phi) (r_2 +s_2 N_\phi) \big)}\Bigg] c_{R+r_1} c_{R+r_2} c_{R-(r_1+r_2)},
\end{equation}
with the identification $c_{m+N_\phi} = c_m$ and
\begin{align}
t_3^0 &= 1 \\
t_3^2 &= \frac{1}{4} H_2 \Big( \frac{\kappa \big( r_1-r_2 + (s_1-s_2) N_\phi \big)}{\sqrt{2}} \Big) + \frac{1}{4} H_2 \Big( \frac{\sqrt{3} \kappa \big( (r_1+r_2) + (s_1+s_2) N_\phi \big)}{\sqrt{2}} \Big) \\
t_3^3 &= \frac{\sqrt{3}}{8} H_2 \Big( \frac{\kappa \big( r_1-r_2 + (s_1-s_2) N_\phi \big)}{\sqrt{2}} \Big) H_1 \Big( \frac{\sqrt{3} \kappa \big( (r_1+r_2) + (s_1+s_2) N_\phi \big)}{\sqrt{2}} \Big) \nonumber \\
&\qquad - \frac{1}{8 \sqrt{3}}  H_3 \Big( \frac{\sqrt{3} \kappa \big( (r_1+r_2) + (s_1+s_2) N_\phi \big)}{\sqrt{2}} \Big).
\end{align}
\end{widetext}
Further pseudopotentials can be obtained by starting from the corresponding expressions of the cylinder geometry. All the pseudopotentials respect many-body translational symmetries in the torus~\cite{chakrabortyFQHEbook, haldane1985many, bernevig2012emergent}.

\subsection{Pseudopotentials in the sphere geometry}
Pseudopotentials in the sphere are projections to the (relative) angular momentum eigenstates in the sphere. Because of the appreciated $SO(3)$ (or $SU(2)$) symmetry, relative angular momentum eigenstates of $N$ particles with relative angular momentum $m$ correspond to $SU(2)$ \textit{spin} $N\frac{N_\phi}{2} - m$ representations of $N$ electrons. Using second quantization language, the angular momentum operators can be written as
\begin{eqnarray}
L_z &=& \sum_{M=-S}^S M c_M^\dagger c_M \nonumber \\
L_+ &=& \sum_{M=-S}^S \sqrt{S(S+1) - M(M+1)} c_{M+1}^\dagger c_M \nonumber \\
L_- &=& \sum_{M=-S}^S \sqrt{S(S+1) - M(M-1)} c_{M-1}^\dagger c_M, \quad
\label{sphere_angular_operators}
\end{eqnarray}
where $c_M$ ($c_M^\dagger$) is a creation (annihilation) operator associated with $\phi_M$ and it satisfies the commutation relation $\big[ c_M, c_{M'}^\dagger \big] = \delta_{M,M'}$. Angular momentum operators satisfy the canonical angular momentum commutation relations:
\begin{eqnarray}
\big[ L_+, L_- \big] &=& 2 L_z \nonumber \\
\big[ L_z, L_\pm \big] &=& \pm L_z \nonumber .
\end{eqnarray}

In the following, we construct pseudopotentials in terms of the second quantization. Unlike the plane geometry, relative angular momentum eigenstates and the pseudopotentials are dependent on the underlying monopole strength $2S = N_\phi$. The two-body pseudopotential is given by
\begin{equation}
P_2^m = \frac{1}{2!} \sum_{M=-J}^J T_{J,M}^\dagger T_{J,M},
\end{equation}
where $J = 2S - m$ and
\begin{equation}
T_{J,M} = \sum_{m_1,m_2 = -S}^S \ip{J,M}{S,M_1;S,M_2} c_{M_1} c_{M_2}.
\end{equation}
$\ip{J,M}{S,M_1;S,M_2}$ is the usual Clebsch-Gordan coefficient of $SU(2)$ representation. $\frac{1}{\sqrt{2!}} T_{J,M}^\dagger$ acting on a vacuum creates a normalized two-particle state with relative angular momentum $M$. Due to the bosonic nature of operators, only even $m$ is allowed (or non-vanishing).

The three-body pseudopotential is given by
\begin{equation}
P_3^{(m,a)} = \frac{1}{3!} \sum_{M=-J}^J T_{J,M}^\dagger T_{J,M},
\end{equation}
where $a$ denotes the possible degeneracy and $J = 3S -m$. The degeneracy (for a sufficiently large $S$) exactly matches with the degeneracy of the (infinite) plane~\cite{simon2007pseudopotentials}. The first three $T_{3S-m,3S-m}$ are as follows:
\begin{widetext}
\begin{eqnarray}
T_{3S,3S} &=& \big(c_S\big)^3 \\
T_{3S-2,3S-2} &=& \sqrt{\frac{12 S}{6S-1}} \big(c_S \big)^2 c_{S-2} - \sqrt{\frac{3(2S-1)}{6S-1}} c_S \big(c_{S-1} \big)^2 \\
T_{3S-3,3S-3} &=& \sqrt{\frac{6 S^2}{(3S-1)(3S-2)}} \big( c_S \big)^2 c_{S-2} - \sqrt{\frac{18 S(S-1)}{(3S-1) (3S-2)}} c_S c_{S-1} c_{S-2} + \sqrt{\frac{2 (S-1)(2S-1)}{(3S-1)(3S-2)}} \big( c_{S-1} \big)^3. \qquad
\end{eqnarray}
\end{widetext}

Given $T_{J,J}$, other operators can be computed using the lowering operator:
\begin{equation}
T_{J,M-1}^\dagger = \frac{1}{\sqrt{J(J+1) - M (M-1)}} \big[L_- ,T_{J,M}^\dagger \big].
\end{equation}

Using the stereographic projection, which maps a sphere to an infinite plane, the annihilation operators are mapped as
\begin{equation}
\textrm{sphere: } c_{m-S} \leftrightarrow \textrm{plane: } c_{m} \nonumber
\end{equation}
and the relative eigenstates in the sphere maps to the relative eigenstates in the sphere:
\begin{equation}
T_{NS-m, -(NS-m) +M}^{(a)} \xrightarrow[S \to \infty]{} \big[ T_N^{(m,a)} \big]_M
\end{equation}
Finally, the Clebsch-Gordan coefficient of the sphere reduces to the Clebsch-Gordan of the plane,
\begin{align}
& \ip{2S-m,-(2S-m)+M}{S,-S+m_1;S,-S+m_2} \nonumber \\
&\qquad \xrightarrow[S \to \infty]{} \ip{M,m}{m_1,m_2},
\label{clebsch_gordan_sphere_plane}
\end{align}
and the similar relations hold for three- and many-particle cases.

\subsection{Pseudopotentials in the coordinate representation}
\label{pseudopotentials_coordinates}
So far, we have expressed operators using only the guiding-center degrees of freedom which capture all the physics in the LLL. On the other hand, it is sometimes useful to find a more familiar coordinate representation of interactions. However, going back to coordinate representation of interactions  requires the knowledge of the action of an interaction in the higher LLs. Two very different potentials can have the same effect in the LLL. With this caveat, we present representatives of the two-body pseudopotentials and the first two of three-body pseudopotentials. We choose the symmetric gauge as we require an explicit gauge in order to specify the action of the operator in the position Hilbert space.

We demand our interaction $V(\r_1,\dots,\r_N)$ to be translationally invariant and symmetric under the exchange of particle indices. The Jacobi coordinate in Eq.~(\ref{Jacobi_coordinates}) is again useful for this purpose as $V$ becomes a function of $\u_1,\dots,\u_{N-1}$ only. It is convenient to work in the momentum space,
\begin{align}
&V(\u_1,\dots,\u_{N-1}) \nonumber \\ 
&\qquad = \prod_{i=1}^{N-1} \bigg( \int \frac{d^2 (l_i \k_i)}{(2\pi)^2} e^{i \k_i \cdot \u_i} \bigg) V(\k_1,\dots,\k_{N-1}). \nonumber
\end{align}
In this section, we share the notation of Ref.~\onlinecite{lee2013pseudopotential}, but our explicit coordinate choice results in some differences. Also, we make our potential as symmetric as possible. The LLL projections and the matrix elements appearing in this section can also be found in Ref.~\onlinecite{murthy2003review}.

Using the LLL projection of the following operator:
\begin{widetext}
\begin{eqnarray}
\overline{e^{i\k_i\cdot\u_i}} &=& \overline{e^{i (\k_i)_x \big(X_i-\frac{l_i^2}{\hbar}(\boldsymbol{\pi}_i)_x \big) + i (\k_i)_y \big(Y_i + \frac{l_i^2}{\hbar} (\boldsymbol{\pi}_i)_y \big)}} = \overline{e^{i \frac{i l_i}{\sqrt{2}} (k_i^* a_i + k_i a_i^\dagger)} } e^{\frac{i l_i}{\sqrt{2}} (k_i^* b_i^\dagger + k_i b_i) } \nonumber \\
&=& e^{-l_i^2 |\k_i|^2 /2} \overline{e^{i \frac{i l_i}{\sqrt{2}} k_i^* a_i} e^{i \frac{i l_i}{\sqrt{2}} k_i a_i^\dagger} } e^{\frac{i l_i}{\sqrt{2}} k_i^* b_i^\dagger} e^{\frac{i l_i}{\sqrt{2}} k_i b_i} \nonumber \\
&=& e^{-l_i^2 |\k_i|^2 /2} e^{\frac{i l_i}{\sqrt{2}} k_i^* b_i^\dagger} e^{\frac{i l_i}{\sqrt{2}} k_i b_i},
\label{LLL_plane_wave}
\end{eqnarray}
the LLL projected $V$ is given by
\begin{equation}
\overline{V}(\u_1,\dots,\u_{N-1}) = \prod_{i=1}^{N-1} \bigg( \int \frac{d^2 (l_i \k_i)}{(2\pi)^2} e^{-l_i^2 |\k_i|^2 /2} \bigg) V(\k_1,\dots,\k_{N-1}) \bigg( \openone_{\mathcal{H}_\textrm{LLL}^\textrm{CM}} \otimes \prod_{i=1}^{N-1} \Big( e^{\frac{i l_i}{\sqrt{2}} k_i^* b_i^\dagger} e^{\frac{i l_i}{\sqrt{2}} k_i b_i} \Big) \bigg).
\end{equation}
From now on, we suppress $\openone_{\mathcal{H}_\textrm{LLL}^\textrm{CM}}$ factor. The matrix element associated with the LLL-projected interaction can be evaluated using (when $m_i \ge m_i'$):
\begin{eqnarray}
\oev{m_i}{e^{\frac{i l_i}{\sqrt{2}} k_i^* b_i^\dagger} e^{\frac{i l_i}{\sqrt{2}} k_i b_i}}{m_i'} &=& \sum_{p=0}^{m_i} \sum_{q=0}^{m'_i} \bra{m_i} \frac{1}{p!} \bigg( \frac{i}{\sqrt{2}} l_i k_i^* \hat{b}_i^\dagger \bigg)^p \frac{1}{q!} \bigg( \frac{i}{\sqrt{2}} l_i k_i \hat{b}_i \bigg)^q \ket{m'_i} \nonumber \\
&=& \sum_{p=m_i-m'_i}^{m_i} \sum_{q=0}^{m'_i} \bra{m_i} \frac{1}{p!} \bigg( \frac{i}{\sqrt{2}} l_i k_i^* \hat{b}_i^\dagger \bigg)^p \frac{1}{q!} \bigg( \frac{i}{\sqrt{2}} l_i k_i \hat{b}_i \bigg)^q \ket{m'_i} \nonumber \\
&=& \bigg( \frac{i}{\sqrt{2}} l_i k_i^* \bigg)^{m_i-m'_i} \sqrt{\frac{m_i!}{m'_i!}} \sum_{p'=0}^{m'_i} \sum_{q=0}^{m'_i} \frac{1}{(p'+(m_i-m'_i))!} \frac{1}{q!} \bra{m_i} \bigg( \frac{i}{\sqrt{2}} l_i k_i^* \hat{b}_i^\dagger \bigg)^{p'} \bigg( \frac{i}{\sqrt{2}} l_i k_i \hat{b}_i \bigg)^q \ket{m'_i} \nonumber \\
&=& \bigg( \frac{i}{\sqrt{2}} l_i k_i^* \bigg)^{m_i-m'_i} \sqrt{\frac{m_i!}{m'_i!}} \sum_{q=0}^{m'_i} \frac{1}{(q+(m_i-m'_i))! q!} \frac{m'_i!}{(m'_i-q)!} \Big( -\frac{l_i^2 |\k_i|^2}{2} \Big)^q \nonumber \\
&=& \bigg( \frac{i}{\sqrt{2}} l_i k_i^* \bigg)^{m_i-m'_i} \sqrt{\frac{m'_i!}{m_i!}} L_{m'_i}^{m_i-m'_i} \Big( \frac{l_i^2 |\k_i|^2}{2} \Big),
\end{eqnarray}
and similarly when $m_i \le m_i'$, 
\begin{equation}
\oev{m_i}{e^{\frac{i l_i}{\sqrt{2}} k_i^* b_i^\dagger} e^{\frac{i l_i}{\sqrt{2}} k_i b_i}}{m_i'} = \bigg( \frac{i}{\sqrt{2}} l_i k_i \bigg)^{m'_i-m_i} \sqrt{\frac{m_i!}{m'_i!}} L_{m_i}^{m'_i-m_i} \Big( \frac{l_i^2 |\k_i|^2}{2} \Big). \nonumber
\end{equation}
The matrix element is given by:
\begin{align}
\oev{m_1,\dots,m_{N-1}}{\overline{V}}{&m'_1,\dots,m'_{N-1}} = \prod_{i=1}^{N-1} \bigg( \int \frac{d^2 (l_i \k_i)}{(2\pi)^2} e^{-l_i^2 |k_i|^2 /2} \bigg) V(\k_1,\dots,\k_{N-1}) \bigg( \prod_{i=1}^{N-1} \oev{m_i}{e^{\frac{i l_i}{\sqrt{2}} k_i^* b_i^\dagger} e^{\frac{i l_i}{\sqrt{2}} k_i b_i} }{m_i'} \bigg) \nonumber \\
&= \prod_{i=1}^{N-1} \bigg( \int \frac{d^2 (l_i \k_i)}{(2\pi)^2} e^{-l_i^2 |k_i|^2 /2} \bigg( \frac{i}{\sqrt{2}} l_i k_i^* \bigg)^{m_i-m'_i} \sqrt{\frac{m'_i!}{m_i!}} L_{m'_i}^{m_i-m'_i} \Big( \frac{l_i^2 |\k_i|^2}{2} \Big) \bigg) V(\k_1,\dots,\k_{N-1}) .
\end{align}
The following identity provides a guiding principle in choosing a representative interaction.
\begin{equation}
\int \frac{d^2 (l \k)}{(2\pi)^2} L_n \Big( \frac{l^2 |\k|^2}{2} \Big) \big( l k \big)^{m'-m} L_m^{m'-m} \Big( \frac{l^2 |\k|^2}{2} \Big) e^{-\frac{1}{2} l^2 |\k|^2} = \frac{1}{2\pi} \delta_{n,m} \delta_{m,m'}
\end{equation}
\end{widetext}
The two-body pseudopotential $P_2^m$ can be represented as 
\begin{equation}
V(\k_1) = 2\pi L_m \bigg( \frac{l_1^2 |\k_1|^2}{2} \bigg),
\end{equation}
where $m$ ($\ge 0$) is an even integer for the bosonic case and an odd integer for the fermionic case. Note that this representation gives translationally invariant potential and symmetric under the exchange of particles, i.e. satisfies $V(-\k_1) = V(\k_1)$ in the momentum space of the Jacobi coordinates.

We demand that the three-body interaction pseudopotentials are translationally invariant interaction, i.e., $V$ is a function of relative coordinates, and remains invariant under the particle exchanges. The latter condition is equivalent to $V(\k_1,\k_2) = V(-\k_1,\k_2) = V(\frac{1}{2} \k_1 + \frac{3}{4} \k_2 , \k_1 - \frac{1}{2} \k_2 )$. In the following, we present the first two three-body pseudopotentials. The higher angular momentum and/or $N$($>3$)-body interactions can also be constructed using similar considerations. The first two relative angular momentum eigenstates of three particles in the relative angular momentum basis $\ket{m_1',m_2'}$ are $\ket{0}_\textrm{rel} = \ket{0,0}$ and $\ket{2}_\textrm{rel} = \frac{1}{\sqrt{2}} \big(\ket{2,0} + \ket{0,2} \big)$. Then a representative expression for $P_3^0$ is 
\begin{equation}
V(\u_1,\u_2) = (2\pi)^2 L_0 \Big( \frac{l_1^2 |\k_1|^2}{2} \Big) L_0 \Big( \frac{l_2^2 |\k_2|^2}{2} \Big),
\end{equation}
and for $P_3^2$ is
\begin{align}
&V(\u_1,\u_2) = (2\pi)^2 \bigg( L_2 \Big( \frac{l_1^2 |\k_1|^2}{2} \Big) L_0 \Big( \frac{l_2^2 |\k_2|^2}{2} \Big) + \nonumber \\
&\,\,\,\, L_1 \Big( \frac{l_1^2 |\k_1|^2}{2} \Big) L_1 \Big( \frac{l_2^2 |\k_2|^2}{2} \Big) + L_0 \Big( \frac{l_1^2 |\k_1|^2}{2} \Big) L_2 \Big( \frac{l_2^2 |\k_2|^2}{2} \Big) \bigg). \nonumber \\
\end{align}
The Gaffnian state~\cite{simon2007construction} is the (highest density) zero-energy ground state of the Hamiltonian $AP_3^0 + BP_3^2$ ($A,B>0$).

\section{Density Operator Algebra}
In the LLL, we quench the dynamical momenta and use only the guiding-center degrees of freedom to describe physics. This yields nontrivial commutation relation among LLL-projected density operators. We follow the modernized definition of the guiding-center (projected) density operator~\cite{haldane2011geometrical, haldane2011selfdual, murthy2003review}.

The density operator and its Fourier transformation are given by $\rho(\r) = \sum_{i=1}^{N_e} \delta^{(2)} (\r-\r_i)$ and $\rho_\k = \sum_{i=1}^{N_e} e^{-i\k \cdot \r_i}$. We are interested in the LLL-projected version of them. Using a similar technique as in Eq.~(\ref{LLL_plane_wave}),
\begin{equation}
\overline{\rho}_\k = \sum_{i=1}^{N_e} \overline{e^{-i\k \cdot \r_i}} = e^{-l_B^2 |\k|^2 /2} \sum_{i=1}^{N_e} e^{-\k \cdot \R_i} \equiv e^{-l_B^2 |\k|^2 /2} \hat\rho_\k, \nonumber
\end{equation}
where we have defined the guiding-center density operator $\hat\rho_\k = \sum_{i=1}^{N_e} e^{-\k \cdot \R_i}$. Note that $\hat\rho_\0 = N_e$ and $\hat\rho_\k^\dagger = \hat\rho_{-\k}$. The LLL projection of a product of two density operators is given by
\begin{align}
\overline{\rho_{\q_1} \rho_{\q_2}} = & e^{-\frac{1}{4} l_B^2 (|\q_1|^2 + |\q_2|^2)} \hat\rho_{\q_1} \hat\rho_{\q_2} \nonumber \\
&+ \bigg( 1- e^{\frac{i}{2} l_B^2 \q_1 \wedge \q_2} e^{\frac{1}{2} l_B^2 \q_1 \cdot \q_2}\bigg) \hat\rho_{\q_1+\q_2}. 
\label{two_density_operator_projection}
\end{align}
Exchanging the role of $\q_1$ and $\q_2$ gives the commutation relation between the density operators which is the GMP algebra~\cite{girvin1986magneto}:
\begin{equation}
[\hat\rho_{\q_1} , \hat\rho_{\q_2}] = 2i \sin \Big( \frac{1}{2} l_B^2 \q_1 \wedge \q_2 \Big) \hat\rho_{\q_1+\q_2}
\label{density_operator_algebra}
\end{equation}
When it comes to a three-body interaction, we need the LLL projection of a product of three density operators. After some tedious algebra, we get
\begin{widetext}
\begin{align}
\overline{\rho_{\q_1+\q_2} \rho_{\q_1}^* \rho_{\q_2}^*} =& e^{-\frac{1}{4} l_B^2 |\q_1+\q_2|^2} e^{-\frac{1}{4} l_B^2 |\q_1|^2} e^{-\frac{1}{4} l_B^2 |\q_2|^2} \hat\rho_{\q_1+\q_2} \hat\rho_{\q_1}^\dagger \hat\rho_{\q_2}^\dagger + \Big(1 - e^{\frac{1}{2} l_B^2 q_1^* q_2} \Big)  e^{-\frac{1}{2} l_B^2 |\q_1+\q_2|^2} \hat\rho_{\q_1+\q_2} \hat\rho_{\q_1+\q_2}^\dagger \nonumber \\ 
&+ \Big(1 - e^{-\frac{1}{2} l_B^2 q_1^* q_2} e^{-\frac{1}{2} l_B^2 |\q_2|^2} \Big) e^{-\frac{1}{2} l_B^2 |\q_1|^2} \hat\rho_{\q_1} \hat\rho_{\q_1}^\dagger + \Big(1 - e^{-\frac{1}{2} l_B^2 q_1 q_2^*} e^{-\frac{1}{2} l_B^2 |\q_1|^2} \Big) e^{-\frac{1}{2} l_B^2 |\q_2|^2}  \hat\rho_{\q_2} \hat\rho_{\q_2}^\dagger \nonumber \\
&+ \bigg( \Big( e^{\frac{1}{2} l_B^2 q_1^* q_2} + e^{\frac{1}{2} l_B^2 q_1 q_2^*} \Big) e^{-\frac{1}{2} l_B^2 |\q_1+\q_2|^2} - e^{-\frac{1}{2} l_B^2 |\q_1+\q_2|^2} - e^{-\frac{1}{2} l_B^2 |\q_1|^2} - e^{-\frac{1}{2} l_B^2 |\q_2|^2} + 1 \bigg) N_e .
\end{align}
\end{widetext}

\section{Density Correlation Functions}
\label{appendix_density_correlation_functions}
Density correlation functions contain useful information of the system in the thermodynamic limit. We will see later that the single-mode approximation is expressed in terms of the density correlations. In this section, we review the density correlation functions - from one-density to three-density correlation function - and discuss how they enter in the expressions for single-mode approximation of a three body interaction. We consider the infinite plane geometry and use the symmetric gauge in this section.

\subsection{Density correlation functions in the thermodynamic limit}
The one-particle density is a density correlation function defined by
\begin{align}
\rho (\r) &= \ev{\hat\psi^\dagger (\r) \hat\psi (\r)} \nonumber \\
&= N_e \int d^2 \r_2 \dots d^2 \r_{N_e} |\Psi(\r,\r_2,\dots,\r_{N_e})|^2,
\end{align}
where $\hat\psi(\r) = \sum_{m} \phi_m (\r) c_m$ is the annihilation operator at $\r$ and $c_m$ is the annihilation operator associated with orbital $m$. We are mainly interested in a homogeneous liquid: $\rho(\r) \to \rho = \frac{\nu}{2\pi}$ in the thermodynamic limit.

The pair-correlation function is a two-density correlation function defined by
\begin{align}
g(\R,&\u) = \rho^{-2} \ev{\hat\psi^\dagger (\r_1) \hat \psi^\dagger (\r_2) \hat \psi(\r_2) \hat \psi(\r_1)} \nonumber \\
&= \frac{N_e (N_e-1)}{\rho^2} \int d^2 \r_3 \dots d^2 \r_{N_e} |\Psi(\r_1,\dots,\r_{N_e})|^2,
\end{align}
where $(\R,\u)$ and $(\u_1,\u_2)$ are related by the two-body Jacobi coordinates: $(\r_1,\r_2) = \big( \R + \frac{\u}{2}, \R - \frac{\u}{2} \big)$. The pair-correlation function remains invariant under the exchange of particle indicies, which gives the relation $g(\R,-\u) = g(\R,\u)$. For a homogeneous  state, $g(\R,\u) \to g(\u)$ in the thermodynamic limit. The symmetry becomes $g(\u) = g(-\u)$ and if we further assume the rotation symmetry, the pair-correlation function becomes rotationally symmetric: $g(\u) = g(|\u|)$.

The three-density correlation function is defined by
\begin{align}
h(\R,\u_1,\u_2&) = \rho^{-3} \ev{\hat\psi^\dagger (\r_1) \hat \psi^\dagger (\r_2) \hat \psi^\dagger (\r_3) \hat \psi(\r_3) \hat \psi(\r_2) \hat \psi(\r_1)} \nonumber \\
&= \frac{N_e (N_e-1) (N_e-2)}{\rho^3} \int d^2 \r_4 \dots d^2 \r_{N_e} |\Psi|^2,
\end{align}
where we have used the three-body Jacobi coordinates: $(\r_1,\r_2,\r_3) = \big(\R+\frac{\u_1}{2}+\frac{\u_2}{3}, \R-\frac{\u_1}{2}+\frac{\u_2}{3}, \R - \frac{2}{3} \u_2 \big)$. The expression remains invariant upon exchanging the indicies in $\r$, which gives the symmetries: $h(\R,\u_1,\u_2) = h(\R,-\u_1,\u_2) = h(\R, \frac{1}{2}\u_1 + \u_2, \frac{3}{4} \u_1 - \frac{1}{2} \u_2 )$. For a homogeneous state, $h(\R,\u_1,\u_2) \to h(\u_1,\u_2)$ in the thermodynamic limit and the symmetries reduce to: 
\begin{equation}
h(\u_1,\u_2) = h(-\u_1,\u_2) = h(\frac{1}{2}\u_1 + \u_2, \frac{3}{4} \u_1 - \frac{1}{2} \u_2 ).
\end{equation}

\subsection{Structure factors in the thermodynamic limit}
The static structure factors are the Fourier transform of the density correlation functions. In defining these structure factors, there exist potential divergences for which we need to regularize. For example, the one-particle density becomes a constant function in the thermodynamic limit (for a homogeneous state), so its Fourier transformation is a Dirac delta function peaked at 0, i.e., divergence at the momentum equals zero. The structure factor is defined as
\begin{equation}
s(\k) = \frac{1}{N_e} \ev{\rho_\k \rho_\k^*} - \rho (2\pi)^2 \delta^{(2)} (\k),
\end{equation}
where the delta function in the second term compensates the divergences of the first term at $\k=0$. The structure factor can be simplified as
\begin{widetext}
\begin{eqnarray}
s(\k) &=& 1 + \frac{N_e (N_e-1)}{N_e} \bigg( \prod_{i=1}^{N_e} \int d^2 \r_i \bigg) e^{i \k \cdot (\r_1 - \r_2)} |\Psi(\r_1,\dots,\r_{N_e})|^2 - \rho (2\pi)^2 \delta^{(2)} (\k) \nonumber \\
&=& 1+ \frac{\rho^2}{N_e} \int d^2 \R \int d^2 \r e^{i\k \cdot \u} g(\r) - \rho \int d^2 \r e^{i \k \cdot \r} \nonumber \\
&=& 1+ \int d^2 \u e^{i \k \cdot \u} \rho \big( g(\u) - 1 \big),
\label{structure_factor_computations}
\end{eqnarray}
where we have used the Jacobi coordinate transform and the fact $\int d^2 \R \to \frac{N_e}{\rho}$ as $N_e \to \infty$. The symmetry of $g$, $g(-\u) = g(\u)$, now translates into the symmetry of $s$ as $s(-\k) = s(\k)$. If we assume the state is rotationally symmetric, then we have: $s(\k) = s(|\k|)$. Since the structure factor vanished at $\k=\0$, it gives the following sum rule:
\begin{equation}
\int d^2 \u e^{i \k \cdot \u} \rho \big( g(\u) - 1 \big) = -1
\end{equation}

The three-point structure factor is defined by
\begin{align}
\Lambda(\k_1,\k_2) = &\frac{1}{N_e} \ev{\rho_{\k_2} \rho_{\k_1+ \frac{1}{2} \k_2}^* \rho_{- \k_1+ \frac{1}{2} \k_2}^*} - s(\k_1) \rho (2\pi)^2 \delta^{(2)} (\k_2) - s(\k_2) \rho (2\pi)^2 \delta^{(2)} \Big(\k_1+ \frac{1}{2} \k_2 \Big) \nonumber \\
&- s(\k_2) \rho (2\pi)^2 \delta^{(2)} \Big(\k_1- \frac{1}{2} \k_2 \Big) - \rho (2\pi)^2 \delta^{(2)} (\k_1) \rho (2\pi)^2 \delta^{(2)} (\k_1),
\end{align}
where all the possible divergences at $\k_2=0$, $\k_1+ \frac{1}{2}\k_2 = 0$, and $\k_1- \frac{1}{2}\k_2 = 0$ are canceled by Dirac delta functions. Using the similar methods in Eq.~(\ref{structure_factor_computations}), the three-point structure factor can be expressed as
\begin{align}
\Lambda(\k_1,\k_2) = &-2  + s\big( \k_1 + \frac{1}{2} \k_2 \big) + s\big( -\k_1 + \frac{1}{2} \k_2 \big) + s(\k_2 ) \nonumber \\
&+ \int d^2 \u_1 \int d^2 \u_2 e^{i\k_1 \cdot \u_1} e^{i\k_2 \cdot \u_2} \rho^2 \Big( \eta(\u_1,\u_2) -g\big(\frac{1}{2}\u_1 + \u_2\big) -g\big(-\frac{1}{2}\u_1 + \u_2\big) -g(\u_1) +2 \Big).
\end{align}
After a bit of work, we can check that the three-point structure factor has the relevant symmetries: 
\begin{equation}
\Lambda(\k_1,\k_2) = \Lambda(-\k_1,\k_2) = \Lambda\Big(\frac{1}{2}\k_1 + \frac{3}{4}\k_2,\k_1-\frac{1}{2}\k_2 \Big).
\label{three_structure_factor_symmetries}
\end{equation}
Since $\Lambda(\k_1,\k_2)$ vanished at $\k_2 = \0$, it gives the following sum rule:
\begin{equation}
\int d^2 \u_2 \, \rho^2 \bigg( h(\u_1,\u_2) - g \Big( \frac{1}{2} \u_1 + \u_2 \Big) - g \Big( -\frac{1}{2} \u_1 + \u_2 \Big) -g(\u_1) +2 \bigg) = -2 \rho \big( g(\u_1) - 1\big).
\end{equation}

\subsection{Guiding-center (projected) structure factors in the thermodynamic limit}
Let's define guiding-center (projected) structure factors and express them in terms of the ordinary structure factors. The projected structure factor is defined by 
\begin{align}
\hat{s}(\k) &= \frac{1}{N_e} \ev{\hat\rho_\k \hat\rho_\k^\dagger} - \rho (2\pi)^2 \delta^{(2)}(\k) \nonumber \\
&= e^{\frac{1}{2}l_B^2 |\k|^2} \Big( \frac{1}{N_e} \ev{\rho_\k \rho_\k^*} - \rho (2\pi)^2 \delta^{(2)}(\k) \Big) - \big(e^{\frac{1}{2} l_B^2 |\k|^2} -1 \big) \nonumber \\
&= e^{\frac{1}{2}l_B^2 |\k|^2} s(\k) - \big(e^{\frac{1}{2} l_B^2 |\k|^2} -1 \big),
\end{align}
where we have used Eq.~(\ref{two_density_operator_projection}) from line 2 to line 3. (The state is assumed to be in the LLL so that the overall projection can be removed.) The projected structure factor has the same symmetry as $s(\k)$: $\hat{s}(-\k) = \hat{s}(\k)$, and when the state is rotationally invariant: $s(\k) = s(|\k|)$.

The guiding-center (projected) three-point structure factor can be expressed as
\begin{align}
\hat{\Lambda} (\k_1,&\k_2) = \frac{1}{N_e} \ev{\hat\rho_{\k_2} \hat\rho_{\k_1+ \frac{1}{2} \k_2}^\dagger \hat\rho_{- \k_1+ \frac{1}{2} \k_2}^\dagger} - \hat{s}(\k_1) \rho (2\pi)^2 \delta^{(2)} (\k_2) - \hat{s}(\k_2) \rho (2\pi)^2 \delta^{(2)} (\k_1+ \frac{1}{2} \k_2) \nonumber \\
&\qquad \quad - \hat{s}(\k_2) \rho (2\pi)^2 \delta^{(2)} (\k_1- \frac{1}{2} \k_2) - \rho (2\pi)^2 \delta^{(2)} (\k_1) \rho (2\pi)^2 \delta^{(2)} (\k_1) \nonumber \\
&= - \Big( e^{\frac{i}{2} l_B^2 \k_1 \wedge \k_2} + e^{-\frac{i}{2} l_B^2 \k_1 \wedge \k_2} \Big) +  e^{\frac{i}{2} l_B^2 \k_1 \wedge \k_2} \hat{s}(\k_2) + e^{-\frac{i}{2} l_B^2 \k_1 \wedge \k_2} \hat{s}\Big( \k_1 + \frac{1}{2}\k_2 \Big) + e^{\frac{i}{2} l_B^2 \k_1 \wedge \k_2} \hat{s}\Big( -\k_1 + \frac{1}{2}\k_2 \Big) \nonumber \\
&+ e^{\frac{1}{4} l_1^2 |\k_1|^2} e^{\frac{1}{4} l_2^2 |\k_2|^2} \int d^2 \u_1 \int d^2 \u_2 e^{i\k_1 \cdot \u_1} e^{i\k_2 \cdot \u_2} \rho^2 \bigg( h(\u_1,\u_2) -g\Big(\frac{1}{2}\u_1 +\u_2\Big) -g\Big(-\frac{1}{2}\u_1 +\u_2\Big) -g(\u_1) +2 \bigg),
\end{align}
where we have used $\k_1 \wedge \k_2 = (\k_1)_x (\k_2)_y - (\k_1)_y (\k_2)_x$. Symmetries are restored by introducing the symmetrized version of three-point structure factor:
\begin{align}
&\hat{\Lambda}^\textrm{sym} (\k_1,\k_2) = \frac{1}{6} \bigg[ \hat{\Lambda}(\k_1,\k_2) + \hat{\Lambda}(-\k_1,\k_2) + \hat{\Lambda}\Big(\frac{1}{2}\k_1 + \frac{3}{4}\k_2,\k_1-\frac{1}{2}\k_2 \Big) + \hat{\Lambda}\Big(-\frac{1}{2}\k_1 + \frac{3}{4}\k_2,-\k_1-\frac{1}{2}\k_2 \Big) \nonumber \\
&\qquad \qquad \qquad \qquad + \hat{\Lambda}\Big(-\frac{1}{2}\k_1 - \frac{3}{4}\k_2,\k_1-\frac{1}{2}\k_2 \Big) + \hat{\Lambda}\Big(\frac{1}{2}\k_1 - \frac{3}{4}\k_2,-\k_1-\frac{1}{2}\k_2 \Big) \bigg] \nonumber \\
&= 2 \cos\Big(\frac{1}{2} l_B^2 \k_1 \wedge \k_2 \Big) \bigg( -1 + \frac{1}{2}\hat{s}(\k_2) + \frac{1}{2}\hat{s}\big( \k_1 + \frac{1}{2}\k_2 \big) + \frac{1}{2}\hat{s}\big( -\k_1 + \frac{1}{2}\k_2 \big) \bigg) \nonumber \\
&\quad + e^{\frac{1}{4} l_1^2 |\k_1|^2} e^{\frac{1}{4} l_2^2 |\k_2|^2} \int d^2 \u_1 d^2 \u_2 e^{i\k_1 \cdot \u_1} e^{i\k_2 \cdot \u_2} \rho^2 \Big( \eta(\u_1,\u_2) -g\big(\frac{1}{2}\u_1 +\u_2\big) -g\big(-\frac{1}{2}\u_1 +\u_2\big) -g(\u_1) +2 \Big).
\end{align}
Which has the following symmetries analogous to Eq.~(\ref{three_structure_factor_symmetries}):
\begin{equation}
\hat\Lambda^{\textrm{sym}}(\k_1,\k_2) = \hat\Lambda^{\textrm{sym}}(-\k_1,\k_2) = \hat\Lambda^{\textrm{sym}}\Big(\frac{1}{2}\k_1+ \frac{3}{4}\k_2,\k_1-\frac{1}{2}\k_2 \Big).
\end{equation}
\end{widetext}

\section{Single-Mode Approximation of the magnetoroton Mode}
\label{appendix_SMA}
In the seminal paper~\cite{girvin1986magneto} by Girvin, MacDonald, and Platzman, the ansatz wave excitation function $\ket{\Psi^\textrm{SMA}_\q} = N_e^{-1/2} \hat\rho_\q^\dagger \ket{\Psi}$ is used to approximate the magnetoroton mode, the low-lying neutral excitation mode. Using this ansatz wavefunction, we gap function can be computed as
\begin{equation}
\Delta(\q) = \frac{ \oev{\Psi_\q^\textrm{SMA}}{(H - E_\textrm{GS})}{\Psi_\q^\textrm{SMA}} }{ \ip{\Psi_\q^\textrm{SMA}}{\Psi_\q^\textrm{SMA}} } = \frac{\hat{f}(\q)}{\hat{s}(\q)},
\label{gap_function}
\end{equation}
where $\hat{s}(\q)$ is the guiding-center structure factor, $V$ is the interaction Hamiltonian, and
\begin{equation}
\hat{f}(\q) = \frac{1}{2 N_e} \oev{\Psi_\q^\textrm{SMA}}{\big[ \hat\rho_\q, \big[\overline{V}, \hat\rho_\q^\dagger\big] \big]}{\Psi_\q^\textrm{SMA}}.
\label{denominator_gap_function}
\end{equation}

In the following, we first review the SMA for two-body interaction and then derive the results for three-body interaction.

\subsection{Two-body interaction}
The interaction potential of two-body interaction in $N_e$ electrons can be written as
\begin{eqnarray}
V(\r_1,\dots,\r_{N_e}) &=& \frac{1}{2} \sum_{i\ne j} V(\r_i-\r_j) \nonumber \\
&=& \frac{1}{2} \int \frac{d^2 (l_1 \k)}{(2\pi)^2} V(\k) (\rho_k \rho_k^* - N_e), \nonumber
\end{eqnarray}
where we used the length scale $l_1 = \sqrt{2}l_B$ of (a Jacobi-coordinate) $\u_1 = \r_1-\r_2$. After the LLL projection, the potential becomes $\overline{V} = \frac{1}{2} \int \frac{d^2 (l_1 \k)}{(2\pi)^2} v(\k) (\hat\rho_k \hat\rho_k^\dagger - N_e)$, where $v(\k) = V(\k) e^{-\frac{1}{4} l_1^2 |\k|^2}$. Using the GMP algebra Eq.~(\ref{density_operator_algebra}), we get
\begin{widetext}
\begin{eqnarray}
[\overline{V}, \hat\rho_\q^\dagger] &=& \frac{1}{2} \int \frac{d^2 (l_1 \k)}{(2\pi)^2} \big( v(\k) - v(\k-\q) \big) 2i \sin \big( \frac{1}{2} l_B^2 \q \wedge \k \big) \hat\rho_{-\q+\k} \hat\rho_{-\k} \nonumber \\
{[}\hat\rho_\q,[\overline{V}, \hat\rho_\q^\dagger]] &=& \frac{1}{2} \int \frac{d^2 (l_1 \k)}{(2\pi)^2} \big( v(\k+\q) - 2 v(\k) + v(\k-\q) \big) 2 \sin^2 \big( \frac{1}{2} l_B^2 \q \wedge \k \big) \hat\rho_\k \hat\rho_{-\k}, \nonumber
\end{eqnarray}
and finally the gap function is given by $\Delta(\q) = \hat{f}(\q)/\hat{s}(\q)$, where 
\begin{equation}
\hat{f}(\q) = \int \frac{d^2 (l_1 \k)}{(2\pi)^2} \Big( v(\k+\q) - 2 v(\k) + v(\k-\q) \Big) 2 \sin^2 \Big( \frac{1}{2} l_B^2 \q \wedge \k \Big) \hat{s}(\k).
\end{equation}

\subsection{Three-body interaction}
We demand that the three-body interaction that appears in the potential is translationally invariant and symmetric under the exchange of particles. Then a three-body interaction becomes $V(\r_1,\r_2,\r_3) = V(\u_1,\u_2)$, where $\u_1$ and $\u_2$ are the relative Jacobi-coordinates in Eq.~(\ref{Jacobi_coordinates}). Particle exchange symmetries are equivalent to $V(\u_1,\u_2) = V(-\u_1,\u_2) = V(\frac{1}{2}\u_1+\u_2,\frac{3}{4}\u_1-\frac{1}{2}\u_2)$. The Fourier transformation of $V(\u_1,\u_2)$ is $V(\u_1,\u_2) = \int \frac{d^2 (l_1 \k_1)}{(2\pi)^2} \int \frac{d^2 (l_2 \k_2)}{(2\pi)^2} V(\k_1,\k_2) e^{i\k_1 \cdot \u_1} e^{i\k_2 \cdot \u_2}$ and the symmetry conditions become $V(\k_1,\k_2) = V(-\k_1,\k_2) = V\big(\frac{1}{2}\k_1+\frac{3}{4}\k_2, \k_1-\frac{1}{2}\k_2 \big)$.

A three-body interaction of $N$ electrons in the coordinate representation is 
\begin{align}
V(\r_1,\dots,\r_N) &= \sum_{1\le i<j<k \le N} V(\r_i,\r_j,\r_k) = \sum_{\substack{i,j,k=1 \\ (i,j,k):\textrm{distinct}}}^{N} V \Big( \r_i-\r_j , \frac{\r_i+\r_j - 2\r_k}{2} \Big) \nonumber \\
&= \frac{1}{6} \int \frac{d^2 (l_1 \k_1)}{(2\pi)^2} \int \frac{d^2 (l_2 \k_2)}{(2\pi)^2} V(\k_1,\k_2) \Big( \rho_{\k_2} \rho^*_{\k_1+\frac{1}{2}\k_2} \rho^*_{-\k_1+\frac{1}{2}\k_2} - \rho_{\k_2} \rho^*_{\k_2} - \rho_{\k_1 + \frac{1}{2}\k_2} \rho^*_{\k_1 + \frac{1}{2}\k_2}  \nonumber \\
& \qquad \qquad \qquad \qquad \qquad \qquad \qquad \qquad \quad +\rho_{-\k_1 + \frac{1}{2}\k_2} \rho^*_{-\k_1 + \frac{1}{2}\k_2} + 2N_e \Big)
\end{align}
and its LLL projection is
\begin{align}
\overline{V} = \frac{1}{6} \int \frac{d^2 (l_1 \k_1)}{(2\pi)^2} \int \frac{d^2 (l_2 \k_2)}{(2\pi)^2} v(\k_1,\k_2) \bigg( &\hat\rho_{\k_2} \hat\rho_{\k_1 + \frac{1}{2}\k_2}^\dagger \hat\rho_{-\k_1 + \frac{1}{2}\k_2}^\dagger - e^{\frac{i}{2} l_B^2 \k_1 \wedge \k_2} \hat\rho_{\k_2} \hat\rho_{\k_2}^\dagger - e^{- \frac{i}{2} l_B^2 \k_1 \wedge \k_2} \hat\rho_{\k_1 + \frac{1}{2}\k_2} \hat\rho_{\k_1 + \frac{1}{2}\k_2}^\dagger \nonumber \\
& - e^{\frac{i}{2} l_B^2 \k_1 \wedge \k_2} \hat\rho_{-\k_1 + \frac{1}{2}\k_2} \hat\rho_{-\k_1 + \frac{1}{2}\k_2}^\dagger + \Big( e^{\frac{i}{2} l_B^2 \k_1 \wedge \k_2} + e^{-\frac{i}{2} l_B^2 \k_1 \wedge \k_2} \Big) N_e \bigg),
\end{align}
where we have introduced $v(\k_1,\k_2) = V(\k_1, \k_2) e^{-\frac{1}{4}l_1^2 |\k_1|^2} e^{-\frac{1}{4}l_2^2 |\k_2|^2}$. Using the symmetries of $v(\k_1,\k_2)$, we can simplify the expression further:
\begin{align}
\overline{V} = &\frac{1}{6} \int \frac{d^2 (l_1 \k_1)}{(2\pi)^2} \int \frac{d^2 (l_2 \k_2)}{(2\pi)^2} v(\k_1, \k_2) \hat\rho_{\k_2} \hat\rho_{\k_1 + \frac{1}{2}\k_2}^\dagger \hat\rho_{-\k_1 + \frac{1}{2}\k_2}^\dagger - \frac{1}{2} \int \frac{d^2 (l_2 \k_2)}{(2\pi)^2} v_2(\k_2) \hat\rho_{\k_2} \hat\rho_{\k_2}^\dagger \nonumber \\
&+ \frac{1}{6} \int \frac{d^2 (l_1 \k_1)}{(2\pi)^2} \int \frac{d^2 (l_2 \k_2)}{(2\pi)^2} v(\k_1, \k_2) \Big( e^{\frac{i}{2} l_B^2 \k_1 \wedge \k_2} + e^{-\frac{i}{2} l_B^2 \k_1 \wedge \k_2} \Big) N_e,
\end{align}
where we have defined $v_2(\k_2) = \int \frac{d^2 (l_1 \k_1)}{(2\pi)^2} v(\k_1, \k_2) e^{\frac{i}{2} l_B^2 \k_1 \wedge \k_2}$. Eq.~(\ref{denominator_gap_function}) for three-body interaction can be computed using the GMP algebra Eq.~(\ref{density_operator_algebra}). After very lengthy algebra, the final expression for the gap function of three-body interaction is
\begin{equation}
\hat{f}(\q) = \int \frac{d^2 (l_1 \k_1)}{(2\pi)^2} \int \frac{d^2 (l_2 \k_2)}{(2\pi)^2} v^{\textrm{sym}}(\k_1,\k_2;\q) \hat{\Lambda}^{\textrm{sym}}(\k_1,\k_2) + \int \frac{d^2 \k}{(2\pi)^2} \Big( 3\rho v_1(\k;\q) - \frac{3}{2} v_2(\k;\q) \Big) \hat{s} (\k),
\end{equation}
where 
\begin{eqnarray}
v^{\textrm{sym}}(\k_1,\k_2;\q) &=& - \Big( v\big(\k_1+\q,\k_2\big) + v\big(\k_1-\q,\k_2\big) \Big) \sin \big(\frac{1}{2} l_B^2 \q \wedge (-\k_1 + \frac{1}{2}\k_2) \big) \sin \big(\frac{1}{2} l_B^2 \q \wedge (\k_1 + \frac{1}{2}\k_2) \big) \nonumber \\
& & -v\big(\k_1,\k_2\big) \sin^2 \big(\frac{1}{2} l_B^2 \q \wedge \k_2 \big) \\
v_1(\k_1;\q) &=& \Big( v \big(\k_1 +\q, \0 \big) -2v\big(\k_1,\0\big) + v \big(\k_1 -\q, \0 \big) \Big) \sin^2 \big(\frac{1}{2} l_B^2 \q \wedge \k_1 \big) \\
v_2(\k_2) &=& \int \frac{d^2 (l_1 \k_1)}{(2\pi)^2} v(\k_1, \k_2) e^{\frac{i}{2}l_B^2 \k_1 \wedge \k_2} \\
v_2(\k_2;\q) &=& \big( v_2(\k_2+\q) - 2v_2(\k_2) + v_2(\k_2-\q) \big) \sin^2 \big( \frac{1}{2} l_B^2 \q \wedge \k_2).
\end{eqnarray}
\end{widetext}


\begin{thebibliography}{99}
\bibitem{FootnoteSMA}
In a plane, the homogeneity of the wavefunction is well-defined only in the thermodynamic limit. In contrast, a state (with finite number of particles) in a sphere is called homogeneous if it has the total angular momentum $0$. Similar criteria hold for a state in a torus.
\end{thebibliography}
\end{document}